\documentclass[letterpaper,twocolumn,10pt]{article} \usepackage{usenix-2020-09}
\usepackage{amsmath,amssymb}
\usepackage{booktabs}
\usepackage{graphicx}
\usepackage{multirow}
\usepackage{tabularx}
\usepackage{longtable}
\usepackage{pdflscape}
\usepackage{array}
\usepackage{enumitem}
\usepackage{tikz}
\usepackage{ragged2e}

\usetikzlibrary{arrows.meta,backgrounds,calc,fit,positioning,shapes.geometric}
\usepackage{balance}
\usepackage{xspace}
\usepackage{microtype}
\usepackage{placeins}
\microtypecontext{spacing=nonfrench}
\definecolor{cOne}{HTML}{3B5BA5}
\definecolor{cTwo}{HTML}{567DB5}
\definecolor{cThree}{HTML}{7A5195}
\definecolor{cFour}{HTML}{A45A7A}
\definecolor{cFive}{HTML}{B56A45}
\definecolor{cSix}{HTML}{238A8D}
\definecolor{cSeven}{HTML}{168A79}
\definecolor{cEight}{HTML}{B54747}
\definecolor{analysisblue}{HTML}{334E8A}
\definecolor{neutral}{HTML}{5B6573}

\newcommand{\study}{\textsc{MechAudit-40}\xspace}
\newcommand{\audit}{\textsc{MechAudit}\xspace}
\newcommand{\Cat}[1]{\textsf{C#1}}

\newcolumntype{Y}{>{\raggedright\arraybackslash}X}
\newcolumntype{C}[1]{>{\centering\arraybackslash}p{#1}}
\newcolumntype{L}[1]{>{\raggedright\arraybackslash}p{#1}}

\hypersetup{
  pdftitle={Measuring Forty LLM Attacks: Cross-Mechanism Internal Signatures and White-Box Auditing},
  pdfauthor={Anonymous submission},
  pdfsubject={Measurement and white-box auditing of LLM attacks}
}

\date{} \title{\Large \bf \study: White-Box Auditing across 40 LLM Attack Mechanisms}
\author{
{\rm Zhen Guo}\\
Saint Louis University\\
zhen.guo.2@slu.edu
\and
{\rm Shanghao Shi}\\
Washington University in St. Louis\\
shanghao@wustl.edu
\and
{\rm Shamim Yazdani}\\
Saint Louis University\\
shamim.yazdani@slu.edu
\and
{\rm Ning Zhang}\\
Washington University in St. Louis\\
zhang.ning@wustl.edu
\and
{\rm Reza Tourani}\\
Saint Louis University\\
reza.tourani@slu.edu
}

\begin{document}
\raggedbottom
\maketitle

\begin{abstract}
While LLM attacks span prompt optimization, multi-turn context manipulation, retrieval poisoning, and model backdoors, white-box defenses are typically evaluated on isolated attack families. Consequently, whether heterogeneous attacks leave internal representation shifts that generalize to unseen threat mechanisms remains unknown. We present \study, a systematic evaluation of 40 attack mechanisms across five open-weight model architectures. Threat-specific success criteria, 100,000 matched clean--attack representation pairs, predefined categories, and grouped holdouts isolate genuine attack-induced displacement from target scale, corpus bias, and data-leakage shortcuts.

Across this testbed, attacks induce structured multi-depth trajectories rather than isolated layer spikes. While raw peaks are non-portable across architectures, target-calibrated profiles preserve transferable geometric signatures: under complete mechanism holdout, hidden states alone recover the threat category of unseen attacks with 82.5\% accuracy. Guided by this finding, we design \audit, a runtime auditor that operates under strict zero-oracle constraints without requiring clean baseline traces or attack metadata. \audit detects 81.1\% of held-out attack executions at a 0.70\% false-positive rate and maintains 78.1\% recall when an entire functional category is withheld. In matched comparisons, \audit is the only detector that avoids mechanism-level coverage collapse, maintaining over 50\% recall across all 40 mechanisms. Internal representations thus support cross-mechanism attack-exposure auditing against calibrated benign references, but decouple from downstream task compromise and parameter integrity.
\end{abstract}
\section{Introduction}
\label{sec:introduction}

White-box auditing for large language models (LLMs) faces a fundamental generalization mismatch. While recent defenses monitor intermediate hidden states to detect adversarial interactions~\cite{zhang2025jbshield,liu2026sentinel,zou2025pishield}, their empirical validation is predominantly confined to isolated attack families (e.g., discrete jailbreaks) or specific model architectures. In production deployments, however, adversaries exploit diverse and structurally distinct attack surfaces, ranging from gradient-optimized adversarial suffixes~\cite{zou2023gcg} and indirect prompt injections~\cite{greshake2023indirect} to retrieval-corpus poisoning~\cite{chen2024agentpoison,zou2025poisonedrag} and backdoor trigger activations~\cite{rando2024universal,xu2024instructions,yan2024vpi}. Because existing detectors are fitted to narrow threat primitives, they suffer severe performance degradation when exposed to previously unseen attack mechanics~\cite{wang2026false}. To systematically address this generalization gap, we formalize the problem of \emph{attack-exposure auditing}: determining whether a single inference execution exhibits representation-level anomalies indicative of adversarial interaction relative to a calibrated benign reference distribution. Consequently, a foundational open challenge is whether a unified representation auditor can maintain reliable detection coverage when complete attack primitives and threat categories are withheld during training.

We investigate this problem in a defensive auditing setting where the evaluator possesses white-box visibility to instrument target-model hidden states during inference. Adversaries operate strictly within the access tiers and manipulation budgets established by their respective source formulations, spanning black-box query access, contextual state manipulation, and training-time poisoning. Our study evaluates cross-mechanism and cross-category generalization across calibrated, open-weight target models. In accordance with standard baseline threat models, our empirical evaluations assume non-adaptive adversaries; analyzing adaptive, monitor-aware evasion strategies remains an orthogonal research direction outside our present claims.

Evaluating cross-mechanism transfer requires a structured sampling frame that spans diverse optimization procedures and intervention surfaces. To prevent over-indexing on standard jailbreak templates, we formalize eight distinct threat categories: objective-driven direct-input optimization and feedback-guided iterative search~\cite{zou2023gcg,chao2023pair}; semantic restructuring and carrier transformation~\cite{zeng2024pap,li2024drattack,yuan2024cipherchat}; multi-turn context manipulation~\cite{russinovich2025crescendo,weng2025fitd}; indirect prompt and agent-channel injection~\cite{greshake2023indirect,liu2023houyi}; persistent retrieval and memory poisoning~\cite{zou2025poisonedrag,chen2024agentpoison}; and model-parameter and training backdoors~\cite{yan2024vpi,rando2024universal,xu2024instructions}. Prior to analyzing behavioral outcomes or intermediate representations, we select five representative mechanisms per category, establishing a balanced registry of 40 mechanisms. This design prevents over-represented attack families from distorting aggregate detection metrics and supports controlled category-holdout evaluations. We emphasize that this testbed is constructed for balanced cross-category benchmarking rather than estimating real-world empirical attack prevalence.

Because the registry mechanisms target disparate security properties, \study establishes threat-specific success criteria across diverse open-weight instruction-tuned architectures. Our empirical testbed leverages an extensive corpus of judged execution traces for behavioral validation alongside matched clean--attack internal representation pairs for white-box auditing. To prevent data leakage and benchmark contrivance, grouped train--test partitions strictly isolate related execution traces, ensuring that prompts, dialogue contexts, and optimization seeds do not co-occur across splits.

Leveraging this controlled testbed, we examine whether adversarial perturbations induce internal representation shifts that generalize across disparate mechanisms. By analyzing normalized layer-wise activation differences between matched clean and attacked executions under strict mechanism holdouts, we find that multi-depth hidden-state trajectories preserve distinct mechanism-associated structural signatures across the attack registry. In leave-one-mechanism-out experiments, layer-wise activation profiles reliably recover the category of unseen mechanisms far above chance expectations. Furthermore, this transferability is distributed across the collective multi-depth trajectory rather than localized to isolated peak layers, whose exact depths and activation magnitudes vary across architectures and attack formulations.

Guided by these representational dynamics, we design \audit, a target-calibrated auditing framework for single-trace runtime inspection. While offline signature discovery leverages matched clean--attack representation pairs to train detector components, online deployment inspects solely the observed runtime execution against a pre-calibrated benign reference distribution. \audit compresses multi-depth hidden states via a fixed random projection~\cite{achlioptas2003random}, feeding a composite scoring rule that couples a cross-mechanism ExtraTrees classifier~\cite{geurts2006extratrees} with a benign-only Mahalanobis distance estimator~\cite{lee2018mahalanobis}. This unified architecture is held constant across all evaluated attack classes, requiring only target-specific benign calibration and threshold selection. At runtime, \audit operates under strict zero-oracle constraints, requiring no matched clean baseline, attack category metadata, or downstream outcome labels.

Across our extensive validation testbed, \audit demonstrates robust generalization to unseen threats, sustaining high detection recall alongside an operational, sub-percent false-positive rate under both complete mechanism- and category-level holdouts. In particular, \audit preserves reliable detection coverage across every mechanism in the evaluation registry; in contrast, prior specialized baselines (e.g., PIShield~\cite{zou2025pishield}, Sentinel~\cite{liu2026sentinel}) fail to generalize beyond their specific training distributions.

\noindent\textbf{Our contributions are:}
\begin{itemize}[
  label=\textbullet,
  leftmargin=1.25em,
  labelsep=0.5em,
  itemsep=1pt,
  topsep=2pt
]
\item \textbf{Mechanism-balanced benchmarking.}
We construct a standardized testbed spanning 40 diverse LLM attack mechanisms across eight categories and five open-weight models, supported by over 190,000 judged executions, 100,000 matched representation pairs, and provenance-tracked train--test partitions.

\item \textbf{Cross-mechanism representation discovery.}
We demonstrate that multi-depth hidden-state trajectories preserve transferable structural signatures under complete mechanism holdouts, enabling accurate category identification for unseen threat primitives.

\item \textbf{Unified runtime auditing framework.}
We develop \audit, a lightweight framework for single-trace runtime inspection that operates under strict zero-oracle constraints without requiring matched clean traces or attack metadata at deployment.

\item \textbf{Operational security boundaries.}
Through hard-negative calibration, cryptographic split verification, and task-outcome analyses, we systematically bound the evidentiary scope of representation auditing, delineating empirical attack exposure from downstream safety alignment bypass and parameter tampering.
\end{itemize}
\section{System and Threat Model} 
\label{sec:threat_model}
\label{sec:taxonomy}

\begin{figure*}[!t]
  \centering
  \includegraphics[width=\textwidth]{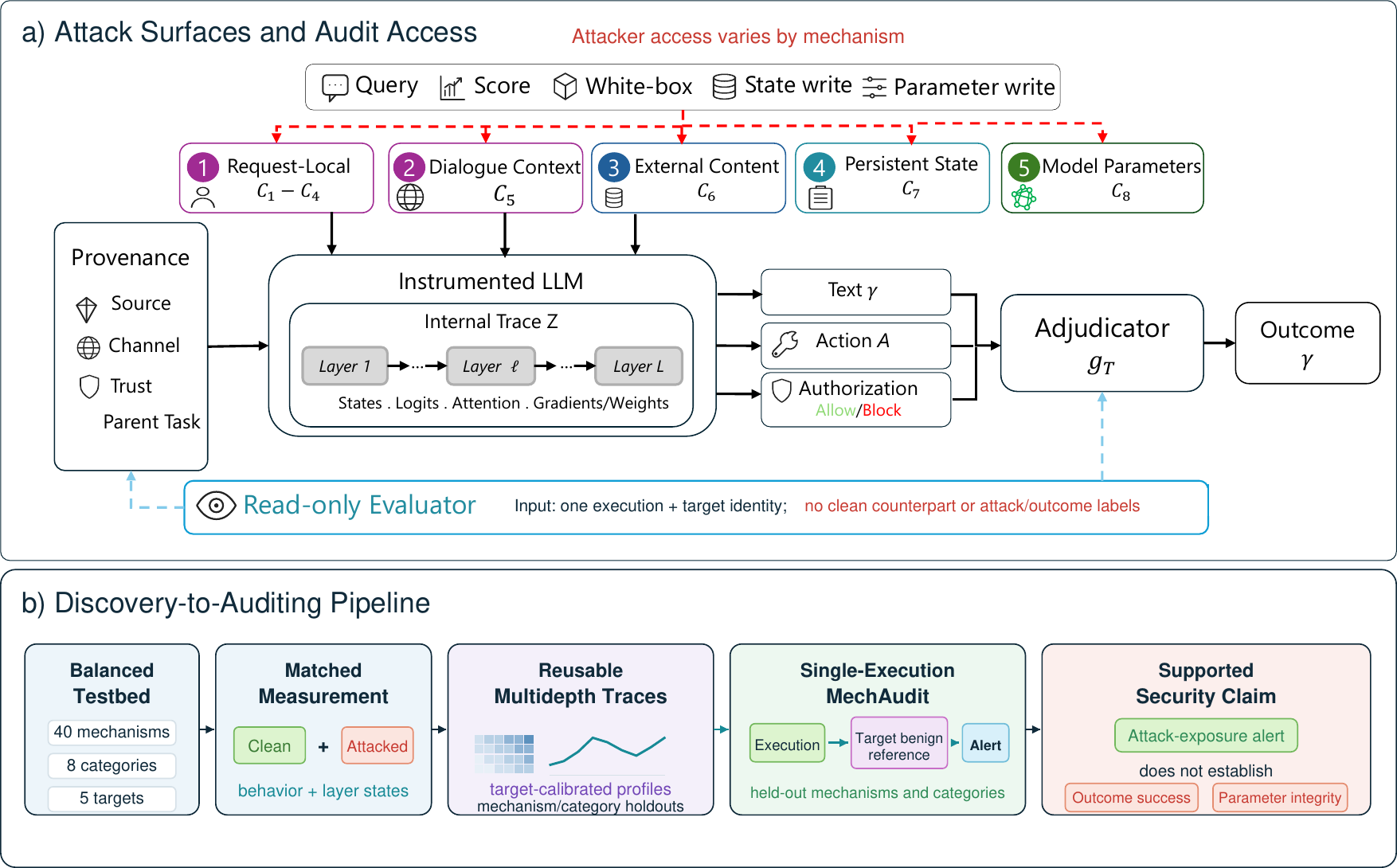}
  \caption{Overview of the \study discovery-to-auditing pipeline. \textbf{(a)} Evaluated threat mechanisms engage target model components under source-specific adversarial access models, while defensive instrumentation captures intermediate layer activations and hidden states. \textbf{(b)} Matched clean--attack execution traces support offline representation discovery. At runtime, \audit scores individual inference traces against an empirical benign reference distribution. Generated alerts provide empirical evidence of adversarial exposure; verifying downstream safety alignment bypass, task compromise, or checkpoint integrity requires orthogonal validation procedures.}
  \label{fig:system-model}
\end{figure*}

To systematically investigate cross-mechanism transfer, we model an LLM application as an end-to-end pipeline spanning system instructions, user prompts, conversational context, retrieval stores, memory records, model parameters, and output actions (Figure~\ref{fig:system-model}). Because adversarial interventions target disparate interfaces across this stack, we evaluate security invariants at the specific component interface where the targeted property is violated rather than at a monolithic text completion~\cite{greshake2023indirect,debenedetti2024agentdojo,chen2024agentpoison}.

\paragraph{Attacker capabilities.}
Adversarial capabilities adhere to the access tiers and budgets defined by each source attack class. Under black-box query access, adversaries interact solely via input interfaces and observe completions~\cite{chao2023pair,mehrotra2024tap}. Under white-box settings, adversaries access weights, gradients, or logits as required by their optimization routines~\cite{guo2024coldattack,zou2023gcg}. For state poisoning, manipulation is strictly bounded to injecting perturbed records into retrieval corpora or memory stores~\cite{chen2024agentpoison,zou2025poisonedrag}. Finally, backdoor adversaries manipulate training data, adapters, or pre-deployment checkpoints~\cite{rando2024universal,yan2024vpi}. Each attack is evaluated in isolation; no adversary commands cumulative capabilities across the registry.

\paragraph{Trusted components and operational regimes.}
The Trusted Computing Base (TCB) comprises the model serving runtime, activation extraction pipeline, execution trace metadata, the trained \audit model, calibrated benign reference representations, and decision thresholds. In accordance with Kerckhoffs's principle, the architectural specifications and algorithms of \audit are public knowledge. We evaluate \audit under two strictly partitioned information regimes: an \textit{offline discovery regime}, where the evaluator leverages attack identities and matched clean--attack state pairs to isolate candidate activation signatures; and an \textit{online deployment regime}, where \audit operates under strict zero-oracle constraints, inspecting a single runtime execution against a calibrated benign reference without access to clean baselines, attack labels, or downstream feedback. Our operational scope is confined to textual executions across open-weight models, excluding closed APIs, multimodal pipelines, and availability attacks.

\paragraph{Mechanism registry and holdouts.}
\label{sec:registry}
We construct our 40-mechanism registry (Table~\ref{tab:taxonomy}) from published LLM attacks satisfying three operational criteria: an algorithmic, reproducible procedure; bounded access requirements; and an empirically measurable safety alignment bypass or task compromise. Multiple prompt templates derived from an identical optimization method map to a single entry, while passive benchmarks and surveys are excluded. To prevent post-hoc bias, each mechanism is assigned to exactly one of eight primary categories based on its intervention surface and optimization primitive prior to hidden-state analysis. This partitioning enforces balanced sampling and supports leave-one-category-out (LOCO) holdout evaluation, where all five mechanisms within a category are completely withheld from training. Appendix~\ref{app:registry} details implementation parameters, perturbation budgets, and exclusion rationales.

\begin{table*}[!t]
  \centering
  \small
  \setlength{\tabcolsep}{6pt}
  \renewcommand{\arraystretch}{0.90}

  \caption{Systematic 40-mechanism LLM attack registry. Each category contains five distinct mechanisms spanning diverse access tiers and intervention surfaces.}
  \label{tab:taxonomy}

  \begin{tabularx}{\textwidth}{
    @{}
    >{\raggedright\arraybackslash}p{0.24\textwidth}
    >{\raggedright\arraybackslash}p{0.45\textwidth}
    >{\raggedright\arraybackslash}X
    @{}
  }
    \toprule

    \textbf{Category} & \textbf{Five Evaluated Mechanisms} & \textbf{Defining Property} \\

    \midrule

    \textbf{C1} Objective-driven direct-input optimization
      & GCG~\cite{zou2023gcg}, BEAST~\cite{sadasivan2024beast}, COLD-Attack~\cite{guo2024coldattack}, ASETF~\cite{wang2024asetf}, AdvPrompter~\cite{paulus2025advprompter}
      & White-box token optimization via gradients, logits, or loss objectives. \\

    \addlinespace[3.5pt]

    \textbf{C2} Behavioral-feedback search
      & PAIR~\cite{chao2023pair}, TAP~\cite{mehrotra2024tap}, AutoDAN~\cite{liu2024autodan}, Rainbow Teaming~\cite{samvelyan2024rainbow}, ECLIPSE~\cite{chen2025eclipse}
      & Iterative black-box prompt search guided by model or judge feedback. \\

    \addlinespace[3.5pt]

    \textbf{C3} Semantic restructuring and persuasion
      & ReNeLLM~\cite{ding2024renellm}, DeepInception~\cite{li2023deepinception}, PAP~\cite{zeng2024pap}, DrAttack~\cite{li2024drattack}, DRA~\cite{liu2024dra}
      & Semantic prompt restructuring while preserving harmful intent. \\

    \addlinespace[3.5pt]

    \textbf{C4} Carrier and representation transformation
      & CipherChat~\cite{yuan2024cipherchat}, ArtPrompt~\cite{jiang2024artprompt}, CodeChameleon~\cite{lv2024codechameleon}, M2S~\cite{ha2025m2s}, FlipAttack~\cite{liu2025flipattack}
      & Syntax, cipher, or encoding shifts to bypass token safety filters. \\

    \addlinespace[3.5pt]

    \textbf{C5} Multi-turn and context-state manipulation
      & Crescendo~\cite{russinovich2025crescendo}, FITD~\cite{weng2025fitd}, DAMON~\cite{zhang2025damon}, RACE~\cite{ying2025race}, PE-CoA~\cite{gao2026pecoa}
      & Contextual state accumulation across multi-turn interactions. \\

    \addlinespace[3.5pt]

    \textbf{C6} Indirect instruction and agent-channel injection
      & Indirect Prompt Injection~\cite{greshake2023indirect}, HouYi~\cite{liu2023houyi}, Adaptive IPI~\cite{schulhoff2025adaptive}, TopicAttack~\cite{wang2025topicattack}, AgentVigil~\cite{wang2025agentvigil}
      & Adversarial instruction injection via external data or tool channels. \\

    \addlinespace[3.5pt]

    \textbf{C7} Persistent retrieval and memory poisoning
      & PoisonedRAG~\cite{zou2025poisonedrag}, AgentPoison~\cite{chen2024agentpoison}, MINJA~\cite{dong2025minja}, Phantom~\cite{chaudhari2024phantom}, Machine Against the RAG~\cite{shafran2025machine}
      & Persistent poisoning of retrieval knowledge bases or memory stores. \\

    \addlinespace[3.5pt]

    \textbf{C8} Model-parameter and training backdoors
      & VPI~\cite{yan2024vpi}, UJ Backdoors~\cite{rando2024universal}, Instr. Backdoors~\cite{xu2024instructions}, BadEdit~\cite{li2024badedit}, BadAgent~\cite{wang2024badagent}
      & Trigger implantation across training corpora, adapters, or weights. \\

    \bottomrule
  \end{tabularx}
\end{table*}

\paragraph{Evaluation objectives.}
\label{sec:outcomes}
Because the 40 mechanisms target distinct pipeline components, attack success cannot be evaluated via a monolithic metric. Instead, success is evaluated by threat-model-specific criteria: safety policy violations for jailbreaks and optimization attacks~\cite{mazeika2024harmbench,chao2024jailbreakbench}; instruction overriding and unauthorized tool execution for indirect prompt injections~\cite{greshake2023indirect,debenedetti2024agentdojo}; output corruption for state poisoning~\cite{zou2025poisonedrag,chen2024agentpoison}; and trigger-conditioned target generation for backdoors~\cite{yan2024vpi,li2025backdoorllm}. Denominators include strictly valid executions, separating benign refusals and infrastructure faults. In contrast, an \audit alert indicates internal representation-level exposure relative to a calibrated benign baseline--not downstream task compromise or checkpoint tampering. Guided by this formulation, our evaluation investigates three research questions: \textbf{RQ1 (Signatures)} examines whether layer-wise hidden-state trajectories encode transferable structural signatures across disparate mechanisms; \textbf{RQ2 (Generalization)} evaluates single-trace auditing efficacy against unseen threats under complete mechanism- and category-level holdouts; and \textbf{RQ3 (Boundaries)} delineates the empirical boundaries of representation auditing across novel architectures, task-level compromises, and parameter integrity.
\section{Measurement Design and Testbed}
\label{sec:testbed}

Empirically answering the research questions formulated in Section~\ref{sec:threat_model} requires an unconfounded, reproducible measurement infrastructure. We operationalize the 40-mechanism registry, threat boundaries, and evaluation objectives into a full factorial execution matrix and a paired internal representation dataset. Throughout our pipeline, adversarial inputs, inference executions, behavioral judgments, and hidden-state traces remain decoupled artifacts linked cryptographically via SHA-256 content hashes. This provenance architecture preserves failed attack attempts within the empirical record and prevents execution ordering or pooled distributional statistics from contaminating out-of-distribution transfer evaluations.

\subsection{Behavioral Evaluation}
\label{sec:attack-instantiation}

\paragraph{Target suite and execution matrix.}
The target suite comprises five frozen open-weight model snapshots spanning diverse architectures and scales: Qwen3-4B, Llama-3.2-3B-Instruct, DeepSeek-R1-Distill-Qwen-7B, Gemma-3-12B-it, and GPT-OSS-20B~\cite{qwen2025qwen3,meta2024llama32,deepseek2025r1card,gemma2025gemma3,openai2025gptoss}. These targets span five distinct model families, parameter scales from 3B to 20B, and 25--49 recorded transformer layers. This architectural variance ensures that discovered representation signatures do not overfit to a specific layer layout, while open weights provide the visibility required to extract layer-wise hidden representations across the residual stream. Text generation uses pinned prompt templates, native precision, and deterministic decoding configurations; exact model revisions and serving controls appear in Appendix~\ref{app:measurement-protocol}.

\begin{table*}[!t]
  \centering
  \small
  \setlength{\tabcolsep}{3.2pt}
  \renewcommand{\arraystretch}{1.04}
  \caption{Behavioral outcomes across 40 mechanisms. Counts are judged/eligible and success/refusal; rates are percentages. Equal-category ASR weights categories equally, and boldface marks column maxima.}
  \label{tab:core-behavioral-summary}
  \begin{tabularx}{\textwidth}{@{}Y C{0.115\textwidth}C{0.120\textwidth}C{0.070\textwidth}C{0.065\textwidth}C{0.090\textwidth}C{0.105\textwidth}C{0.125\textwidth}@{}}
    \toprule
    Target model & Judged / eligible &
    Success / refusal & Micro ASR &
    Refusal & Equal-cat. ASR &
    Worst-cat. ASR & Max within-cat. range \\
    \midrule
    Qwen3-4B & 38,500 / 34,356 & 13,546 / 2,887 & 39.4 & 7.5 & 39.3 & 52.1 (C5) & \textbf{85.6 pp (C8)} \\
    Llama-3.2-3B & 38,500 / 34,356 & 15,756 / 3,853 & 45.9 & 10.0 & 45.1 & 62.9 (C3) & 47.6 pp (C8) \\
    DeepSeek-R1-7B & 38,500 / 34,356 & \textbf{16,728} / 631 & \textbf{48.7} & 1.6 & \textbf{48.8} & \textbf{63.4 (C3)} & 81.6 pp (C8) \\
    Gemma-3-12B-it & 38,500 / 34,356 & 9,021 / \textbf{9,665} & 26.3 & \textbf{25.1} & 25.7 & 51.3 (C3) & 25.8 pp (C8) \\
    GPT-OSS-20B & 38,500 / 34,356 & 6,289 / 8,052 & 18.3 & 20.9 & 17.9 & 37.5 (C7) & 36.7 pp (C8) \\
    \bottomrule
  \end{tabularx}
\end{table*}

Rather than retaining only attack--model combinations that instantiate cleanly, our execution matrix fully crosses all 40 mechanisms with all five target models. This complete factorial design eliminates survivorship bias, ensuring that target sensitivities, failed prompt constructions, and generation degradations are captured as measured empirical outcomes rather than filtered out. Standard attack--model cells contain 1,000 judged executions, while trigger-matched cells contain 500, yielding a behavioral corpus of 192,500 executions. Each cell executes frozen base tasks under the specific access assumptions defined by its source threat model. Refusals, ill-formed outputs, and runtime faults remain in the empirical record rather than being discarded or iteratively resampled. For backdoor mechanisms, we maintain frozen trigger-present and trigger-absent inputs, evaluating parameter fidelity independently from prompt-level exposure.

\paragraph{Outcome measurement and paired corpus.}
Each execution is evaluated against the threat-specific protected outcome formalized in Section~\ref{sec:outcomes}. Deterministic evaluators inspect program state when the task exposes an explicit authorization decision, retrieval target, tool invocation, or structured ground-truth label. Unstructured conversational responses are judged using a frozen Llama-Guard-3-8B model~\cite{grattafiori2024llama3} alongside a dedicated refusal classifier; backdoor adapter fidelity is evaluated via clean accuracy, attack accuracy, and target-label flip rates. Across all settings, Attack Success Rate (ASR) is computed as successful attempts over valid applicable cases, while the refusal rate measures refusals over total judged interactions. These automated labels serve as measurement instruments rather than absolute semantic ground truth. Every recorded rate retains its numerator, denominator, model identifier, attack mechanism, and judge provenance.

To measure internal representational displacement without task-level confounding, we pair 500 attacked executions with clean executions sharing identical base tasks for every attack--model cell. Holding the underlying semantic task constant isolates the internal displacement induced strictly by the adversarial intervention--including any carrier encoding or contextual manipulation. Matched pairs are established prior to representation extraction, and failed attack attempts are retained in the paired corpus to avoid conditioning representation analysis on known downstream success. Across the 40 mechanisms and five target models, this yields 100,000 exactly matched clean--attack representation pairs.

\begingroup
\small
\begin{table*}[!t]
\centering
\caption{Complete behavioral matrix. Cells report \textbf{ASR / refusal} (\%); $n_{\mathrm{judged}}$ is 1,000 except for A37--A39 (500). Boldface marks each row's highest ASR.}
\label{tab:behavioral-atlas-values}

\small
\begingroup

\setlength{\tabcolsep}{2.0pt}
\renewcommand{\arraystretch}{0.82}

\setlength{\aboverulesep}{0.25ex}
\setlength{\belowrulesep}{0.25ex}
\setlength{\abovetopsep}{0pt}
\setlength{\belowbottomsep}{0pt}

\begin{tabular}{@{}
C{0.038\textwidth}
C{0.038\textwidth}
l
C{0.068\textwidth}
*{5}{C{0.084\textwidth}}
C{0.088\textwidth}
@{}}
\toprule

ID &
Cat. &
Attack mechanism and source &
Judged $n$ &
Qwen3-4B &
Llama-3.2-3B &
DeepSeek-R1-7B &
Gemma-3-12B &
GPT-OSS-20B &
Mean \\

\midrule

A01 & C1 & GCG~\cite{zou2023gcg} & 1,000 & 31.5 / 16.1 & 46.3 / 18.8 & \textbf{51.5} / 0.5 & 21.7 / 22.0 & 12.8 / 23.5 & 32.8 / 16.2 \\
A02 & C1 & BEAST~\cite{sadasivan2024beast} & 1,000 & 34.5 / 23.3 & 47.1 / 16.9 & \textbf{52.8} / 2.5 & 21.1 / 31.8 & 19.6 / 14.3 & 35.0 / 17.8 \\
A03 & C1 & COLD-Attack~\cite{guo2024coldattack} & 1,000 & 42.8 / 10.1 & \textbf{59.8} / 13.1 & 59.6 / 0.7 & 32.1 / 22.6 & 16.4 / 22.4 & 42.1 / 13.8 \\
A04 & C1 & ASETF~\cite{wang2024asetf} & 1,000 & 28.8 / 18.4 & 47.2 / 17.1 & \textbf{48.2} / 1.9 & 15.4 / 43.3 & 16.0 / 29.5 & 31.1 / 22.0 \\
A05 & C1 & AdvPrompter~\cite{paulus2025advprompter} & 1,000 & 56.1 / 4.7 & 55.7 / 13.0 & \textbf{59.0} / 0.3 & 25.8 / 20.6 & 19.3 / 13.6 & 43.2 / 10.4 \\

A06 & C2 & PAIR~\cite{chao2023pair} & 1,000 & 49.3 / 3.5 & \textbf{50.3} / 5.8 & 50.2 / 0.4 & 24.8 / 16.6 & 10.9 / 21.8 & 37.1 / 9.6 \\
A07 & C2 & TAP~\cite{mehrotra2024tap} & 1,000 & 38.6 / 4.9 & \textbf{43.7} / 3.9 & 27.9 / 0.9 & 29.2 / 32.6 & 11.7 / 17.8 & 30.2 / 12.0 \\
A08 & C2 & AutoDAN~\cite{liu2024autodan} & 1,000 & \textbf{62.4} / 2.4 & 57.4 / 5.9 & 50.3 / 0.6 & 19.4 / 44.5 & 8.4 / 16.6 & 39.6 / 14.0 \\
A09 & C2 & Rainbow Teaming~\cite{samvelyan2024rainbow} & 1,000 & 51.7 / 2.2 & \textbf{51.8} / 6.7 & 36.4 / 0.1 & 30.9 / 29.3 & 6.2 / 28.5 & 35.4 / 13.4 \\
A10 & C2 & ECLIPSE~\cite{chen2025eclipse} & 1,000 & 33.3 / 6.4 & \textbf{45.9} / 5.2 & 40.3 / 1.3 & 30.9 / 25.4 & 13.9 / 17.4 & 32.9 / 11.1 \\

A11 & C3 & ReNeLLM~\cite{ding2024renellm} & 1,000 & 55.4 / 5.2 & \textbf{60.9} / 3.5 & 60.8 / 0.2 & 44.4 / 4.7 & 18.8 / 31.1 & 48.1 / 8.9 \\
A12 & C3 & DeepInception~\cite{li2023deepinception} & 1,000 & 48.5 / 6.7 & 55.6 / 5.6 & \textbf{61.4} / 0.1 & 47.1 / 5.2 & 18.0 / 30.0 & 46.1 / 9.5 \\
A13 & C3 & PAP~\cite{zeng2024pap} & 1,000 & 53.4 / 3.3 & \textbf{66.2} / 2.0 & 65.4 / 0.4 & 58.6 / 4.2 & 21.6 / 28.2 & 53.0 / 7.6 \\
A14 & C3 & DrAttack~\cite{li2024drattack} & 1,000 & 50.2 / 4.4 & \textbf{67.7} / 2.6 & 65.9 / 0.6 & 58.7 / 1.4 & 15.9 / 33.5 & 51.7 / 8.5 \\
A15 & C3 & DRA~\cite{liu2024dra} & 1,000 & 47.6 / 5.7 & \textbf{63.9} / 0.9 & 63.7 / 0.6 & 48.0 / 4.2 & 21.3 / 30.9 & 48.9 / 8.5 \\

A16 & C4 & CipherChat~\cite{yuan2024cipherchat} & 1,000 & 16.0 / 12.7 & 27.8 / 8.5 & \textbf{37.8} / 0.2 & 11.7 / 38.8 & 6.5 / 19.0 & 20.0 / 15.8 \\
A17 & C4 & ArtPrompt~\cite{jiang2024artprompt} & 1,000 & 21.3 / 6.6 & 28.6 / 8.4 & \textbf{41.2} / 0.2 & 18.2 / 15.3 & 5.9 / 20.8 & 23.0 / 10.3 \\
A18 & C4 & CodeChameleon~\cite{lv2024codechameleon} & 1,000 & 15.9 / 13.7 & 34.9 / 10.7 & \textbf{40.3} / 0.2 & 16.8 / 32.0 & 6.5 / 17.2 & 22.9 / 14.8 \\
A19 & C4 & M2S~\cite{ha2025m2s} & 1,000 & 15.8 / 13.9 & \textbf{40.2} / 4.9 & 33.1 / 0.2 & 20.9 / 23.5 & 6.5 / 17.9 & 23.3 / 12.1 \\
A20 & C4 & FlipAttack~\cite{liu2025flipattack} & 1,000 & 19.8 / 11.4 & 33.9 / 7.1 & \textbf{39.0} / 0.1 & 14.8 / 41.2 & 4.4 / 13.1 & 22.4 / 14.6 \\

A21 & C5 & Crescendo~\cite{russinovich2025crescendo} & 1,000 & 52.9 / 6.0 & \textbf{64.4} / 9.5 & 55.6 / 1.6 & 34.7 / 27.9 & 23.0 / 15.5 & 46.1 / 12.1 \\
A22 & C5 & Foot-In-The-Door~\cite{weng2025fitd} & 1,000 & 53.7 / 5.1 & \textbf{77.3} / 3.3 & 52.7 / 0.7 & 49.0 / 15.8 & 20.7 / 13.6 & 50.7 / 7.7 \\
A23 & C5 & DAMON~\cite{zhang2025damon} & 1,000 & 50.1 / 9.0 & 55.0 / 13.0 & \textbf{56.6} / 1.3 & 50.6 / 12.5 & 23.3 / 14.5 & 47.1 / 10.1 \\
A24 & C5 & RACE~\cite{ying2025race} & 1,000 & 46.7 / 5.6 & \textbf{50.2} / 12.2 & 49.7 / 1.9 & 47.3 / 14.0 & 22.4 / 15.4 & 43.3 / 9.8 \\
A25 & C5 & PE-CoA~\cite{gao2026pecoa} & 1,000 & 56.9 / 6.4 & \textbf{64.1} / 9.8 & 55.2 / 1.1 & 30.3 / 22.4 & 22.7 / 20.9 & 45.8 / 12.1 \\

A26 & C6 & Indirect Prompt Injection~\cite{greshake2023indirect} & 1,000 & 25.3 / 7.3 & 33.7 / 10.3 & \textbf{40.8} / 0.2 & 13.5 / 32.7 & 13.4 / 32.3 & 25.3 / 16.6 \\
A27 & C6 & HouYi~\cite{liu2023houyi} & 1,000 & 29.7 / 4.3 & 33.3 / 12.1 & \textbf{38.7} / 0.0 & 13.9 / 29.3 & 12.6 / 43.1 & 25.6 / 17.8 \\
A28 & C6 & Adaptive IPI~\cite{schulhoff2025adaptive} & 1,000 & 30.7 / 3.4 & 26.5 / 10.5 & \textbf{35.0} / 0.1 & 13.0 / 26.1 & 11.9 / 28.3 & 23.4 / 13.7 \\
A29 & C6 & TopicAttack~\cite{wang2025topicattack} & 1,000 & 33.3 / 2.5 & 30.4 / 12.7 & \textbf{39.5} / 0.1 & 13.2 / 42.7 & 15.1 / 16.7 & 26.3 / 14.9 \\
A30 & C6 & AgentVigil~\cite{wang2025agentvigil} & 1,000 & 30.0 / 3.8 & 29.2 / 10.0 & \textbf{40.4} / 0.1 & 14.4 / 36.4 & 21.5 / 22.0 & 27.1 / 14.5 \\

A31 & C7 & PoisonedRAG~\cite{zou2025poisonedrag} & 1,000 & 40.8 / 4.6 & 41.4 / 3.3 & \textbf{46.2} / 10.1 & 21.1 / 15.2 & 27.6 / 2.5 & 35.4 / 7.1 \\
A32 & C7 & AgentPoison~\cite{chen2024agentpoison} & 1,000 & 43.1 / 4.9 & 48.2 / 5.4 & \textbf{51.5} / 4.7 & 24.1 / 18.0 & 38.7 / 4.4 & 41.1 / 7.5 \\
A33 & C7 & MINJA~\cite{dong2025minja} & 1,000 & 45.2 / 1.4 & 44.6 / 3.6 & \textbf{47.3} / 6.6 & 19.3 / 12.7 & 42.0 / 6.3 & 39.7 / 6.1 \\
A34 & C7 & Phantom~\cite{chaudhari2024phantom} & 1,000 & 40.1 / 1.5 & 39.2 / 8.4 & \textbf{49.2} / 5.2 & 21.2 / 21.3 & 41.6 / 3.6 & 38.3 / 8.0 \\
A35 & C7 & Machine Against the RAG~\cite{shafran2025machine} & 1,000 & 35.8 / 5.1 & 39.0 / 5.0 & \textbf{44.6} / 5.6 & 18.5 / 31.6 & 37.4 / 3.0 & 35.1 / 10.1 \\

A36 & C8 & Virtual Prompt Injection~\cite{yan2024vpi} & 1,000 & 39.3 / 6.1 & 27.1 / 33.3 & \textbf{53.8} / 2.2 & 6.4 / 53.7 & 24.5 / 36.6 & 30.2 / 26.4 \\
A37 & C8 & Universal JB Backdoors~\cite{rando2024universal} & 500 & 20.2 / 42.0 & 23.2 / 68.6 & \textbf{81.8} / 5.6 & 26.2 / 39.6 & 0.6 / 99.0 & 30.4 / 51.0 \\
A38 & C8 & Instructions as Backdoors~\cite{xu2024instructions} & 500 & \textbf{87.0} / 8.2 & 48.0 / 30.6 & 78.4 / 4.6 & 2.6 / 69.4 & 18.8 / 44.6 & 47.0 / 31.5 \\
A39 & C8 & BadEdit~\cite{li2024badedit} & 500 & \textbf{1.4} / 0.4 & 0.4 / 10.8 & 0.2 / 3.6 & 0.4 / 5.6 & \textbf{1.4} / 0.6 & 0.8 / 4.2 \\
A40 & C8 & BadAgent~\cite{wang2024badagent} & 1,000 & 37.6 / 10.8 & 45.5 / 7.3 & \textbf{51.7} / 2.7 & 19.0 / 37.7 & 37.3 / 7.3 & 38.2 / 13.2 \\

\bottomrule
\end{tabular}

\endgroup
\end{table*}
\endgroup

\subsection{Representation Extraction}
\label{sec:layerwise-states}

\paragraph{Residual stream extraction.}
For each matched pair, the instrumentation pipeline extracts the residual stream activation vector corresponding to the final prompt token across all transformer layers. Extracting representations at this pre-generation boundary standardizes measurement across all threat classes without conditioning internal states on output length or generated tokens. Let $h^a_{i\ell}$ and $h^c_{i\ell}$ denote the attacked and clean hidden states for pair $i$ at layer $\ell$. We measure representational displacement using two complementary metrics: cosine distance isolates directional angular shift independent of activation scale, while relative $\ell_2$ distance captures magnitude displacement normalized by the clean activation norm:
\begin{equation}
 \begin{aligned}
 d_{i\ell}^{\cos} &= 1-
 \frac{\langle h^{a}_{i\ell},h^{c}_{i\ell}\rangle}
 {\lVert h^{a}_{i\ell}\rVert_2\lVert h^{c}_{i\ell}\rVert_2},\\
 d_{i\ell}^{\mathrm{rel}} &=
 \frac{\lVert h^{a}_{i\ell}-h^{c}_{i\ell}\rVert_2}
 {\lVert h^{c}_{i\ell}\rVert_2+\epsilon}.
 \end{aligned}
 \label{eq:paired-state-distances}
\end{equation}
Evaluating both metrics verifies whether depth-wise activation trajectories remain invariant to the choice of distance geometry. Because the target models possess varying layer depths ($L \in [25, 49]$), absolute layer indices are not directly comparable. We normalize layer depth as $u = \ell / (L - 1) \in [0, 1]$ to align relative trajectories across architectures without assuming uniform functional roles across layers. Original high-dimensional hidden states and input hashes are preserved in raw storage, while resampled depth trajectories support aggregate cross-model analysis.

\paragraph{Experimental controls and split integrity.}
Our experimental controls systematically eliminate four confounding shortcuts that could artificially inflate cross-mechanism transfer: record leakage, architecture-specific activation scales, unequal sample counts, and category composition bias. Because the attack mechanism represents the primary unit of transfer, mechanism and category holdout splits strictly isolate all records derived from the held-out units prior to feature normalization or detector fitting. Base-task identifiers, model architectures, and input prompts are grouped into non-overlapping partitions, following established benchmark standards for out-of-distribution evaluation~\cite{gulrajani2021domainbed,koh2021wilds}. Cross-model comparisons enforce identical 500 base task IDs across all mechanisms and targets. Furthermore, behavioral judgments and hidden-state records are joined strictly via cryptographic content hashes, pruning unmatched records rather than relying on sequential row alignment.

Normalization and scaling parameters are computed strictly on the training partitions prescribed by each holdout split. Architecture-specific activation magnitudes are standardized against target-specific benign reference banks. To prevent over-represented attacks from dominating aggregate metrics, we apply per-cell macro-averaging across attack--model combinations, evaluate statistical significance via balanced label permutations, and compute confidence intervals by bootstrapping across cells. Consequently, any observed transfer effect must survive rigorous task grouping, target normalization, balanced weighting, and complete category holdouts. Exact software revisions, hardware environments, and execution checksums are detailed in Appendix~\ref{app:measurement-protocol} and Figure~\ref{fig:measurement-pipeline}.
\section{Discovering Reusable Audit Traces}
\label{sec:within-category}

Leveraging the matched clean--attack testbed established in Section~\ref{sec:testbed}, we investigate \textbf{RQ1}: whether adversarial perturbations induce reusable, mechanism-associated representation shifts across disparate threat classes. Answering this question requires first characterizing the geometric structure of internal displacements across model depth, and subsequently evaluating whether these profiles transfer to completely unseen attack primitives under leave-one-mechanism-out holdouts. Every attack--model cell is weighted equally throughout this evaluation to prevent high-resource attack classes from dominating aggregate findings.

\begin{figure*}[!t]
  \centering
  \includegraphics[width=0.96\textwidth]{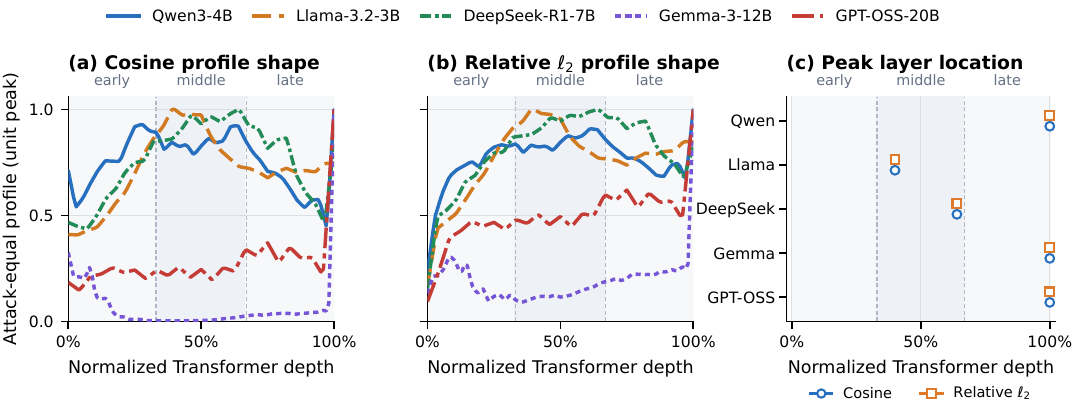}
  \caption{Layer-wise displacement over 100,000 matched clean--attack pairs. \textbf{(a,b)} Attack-equal profiles normalized within target. \textbf{(c)} Peak-depth medians and interquartile ranges. Structured depth profiles lack an absolute layer shared across targets.}
  \label{fig:internal-state-profiles}
\end{figure*}

\subsection{Layer-Wise Profile Geometry}
\label{sec:within-representation}

\paragraph{Decoupling internal state from behavioral outcome.}
Behavioral outcomes contextualize internal state changes rather than serving as ground-truth auditing labels. Across our 40-by-five execution matrix (Table~\ref{tab:core-behavioral-summary} and Table~\ref{tab:behavioral-atlas-values}), Attack Success Rates vary widely (18.3\% on GPT-OSS-20B to 48.7\% on DeepSeek-R1-7B), with within-category ASR ranges reaching 85.6 percentage points. Crucially, internal representational displacement decouples from behavioral success: substantial displacement can accompany faithful attacks, broad parameter damage, or benign carrier shifts, whereas modest displacement can accompany failed prompt instantiations. For instance, the A38 instruction backdoor~\cite{xu2024instructions} maintains >96\% clean accuracy on Qwen, Llama, and Gemma, but collapses to 50.8\% on GPT-OSS due to parameter corruption. To prevent survivorship bias, our representation analysis avoids conditioning on behavioral success, treating internal displacement as evidence of representational exposure rather than downstream exploit completion.

\paragraph{Distributed depth trajectories vs. non-universal peaks.}
Attacks across all eight categories generate structured displacement across model depth rather than isolated spikes at individual layers (Figure~\ref{fig:internal-state-profiles}). Cosine distance and relative-$\ell_2$ displacement exhibit strong directional agreement ($\rho = 0.96$ across 200 attack--model cells), demonstrating that depth-wise trajectories reflect underlying geometric reorganization rather than metric-specific artifacts.

Internal shifts do not concentrate at a universally shared layer. Llama and DeepSeek exhibit broad mid-to-late displacement profiles with median cosine peaks at normalized depths $u = 0.46$ and $0.62$, whereas Qwen, Gemma, and GPT-OSS peak near terminal layers ($u \in [0.91, 0.95]$). Pairwise cross-target rank correlations of raw peak depths remain weak to moderate ($\rho \in [0.03, 0.71]$), ruling out static, portable audit layers. Compact triggers (e.g., A37, A39) introduce prominent tokenization artifacts~\cite{rando2024universal,li2024badedit}, confirming that reusable signatures cannot rely on localized magnitude spikes. Instead, generalizable signatures must capture the relative trajectory across model depth following target-specific calibration.

\paragraph{Geometric profile parameterization.}
To quantify candidate trajectories without relying on raw high-dimensional activations, we summarize each distance profile $q \in \{\cos, \mathrm{rel}\}$ via four geometric descriptors: mean displacement ($\bar d$), peak magnitude ($d^{\max}$), normalized peak depth ($u^{\max}$), and distribution centroid ($c$):
\begin{equation}
  \begin{aligned}
  \mathbf{s}^{q}_{m,a}
    &= [\bar d^{q}_{m,a}, d^{q,\max}_{m,a},
        u^{q,\max}_{m,a}, c^{q}_{m,a}],\\
  \mathbf{z}^{\mathrm{state}}_{m,a}
    &= [\mathbf{s}^{\cos}_{m,a},
        \mathbf{s}^{\mathrm{rel}}_{m,a}].
  \end{aligned}
  \label{eq:category-profile}
\end{equation}
This yields an eight-dimensional state vector per target model, augmented by ASR and refusal rates to form a ten-dimensional joint profile. Features are standardized within each target using strictly the mechanisms permitted by the training split. This representation serves exclusively as a diagnostic probe for signature discovery rather than the runtime auditor.

\subsection{Cross-Mechanism Transferability}
\label{sec:cross-category}

\paragraph{Leave-one-mechanism-out evaluation.}
A consistent average profile does not guarantee out-of-distribution generalization. We therefore evaluate leave-one-mechanism-out (LOMO) category recovery: in each fold, the target mechanism is completely withheld prior to feature standardization and centroid fitting. The held-out mechanism is assigned to the nearest training centroid, with statistical significance established via 5,000 balanced label permutations preserving category sizes.

\begin{table}[!t]
  \centering
  \small
  \setlength{\tabcolsep}{3.0pt}
  \renewcommand{\arraystretch}{1.08}
  \caption{Cross-target structure summary. Appendix Table~\ref{tab:full-crossmodel-structure} reports the complete decomposition.}
  \label{tab:core-crossmodel-structure}
  \begin{tabularx}{\columnwidth}{@{}L{0.25\columnwidth}L{0.35\columnwidth}Y@{}}
    \toprule
    Test & Decisive estimate & Implication \\
    \midrule
    Peak magnitude & \textbf{77.6--86.5\%} attack & Attack-specific magnitude. \\
    Peak depth & \textbf{47.0\%} model; \textbf{40.2\%} attack & No portable layer. \\
    Calibrated depth & \textbf{69.7--78.6\%} attack & Within-target geometry. \\
    Taxonomy holdout & \textbf{82.5--85.0\%} vs. q95 $\leq25.0\%$ & Recoverable clusters. \\
    Target ordering & mean $\rho$ \textbf{0.21--0.55} & Target residuals. \\
    \bottomrule
  \end{tabularx}
\end{table}

Target calibration effectively isolates mechanism-associated structural patterns from architecture-specific scaling (Table~\ref{tab:core-crossmodel-structure}). While raw peak magnitudes are predominantly attack-driven (explaining 77.6--86.5\% of variance), raw cosine peak depth is heavily confounded by model architecture (47.0\% model vs. 40.2\% attack variance). Following target calibration, attack mechanism identity accounts for 78.6\% of cosine peak-depth variation and 69.7\% of relative-$\ell_2$ variation.

\begin{figure}[!t]
  \centering
  \includegraphics[width=\columnwidth]{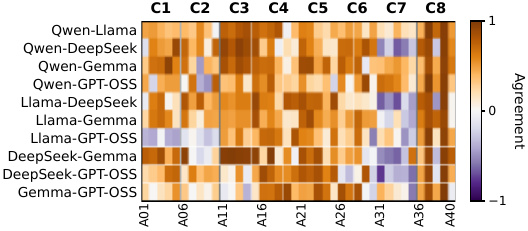}
  \caption{Cross-target agreement of calibrated mechanism profiles. Within-category variation shows that transfer is structured but nonuniform.}
  \label{fig:pairwise-mechanism-fingerprints}
\end{figure}

Calibrated profiles preserve structured yet non-uniform target agreement (Figure~\ref{fig:pairwise-mechanism-fingerprints}), with pairwise rank correlations spanning $\rho = 0.55$ (Llama--Gemma) to $\rho = 0.21$ (DeepSeek--GPT-OSS). Target calibration aligns relative layer depth without forcing artificial uniformity, allowing distinct architectures to preserve characteristic representational geometry.

\begin{figure}[!t]
  \centering
  \includegraphics[width=0.74\columnwidth]{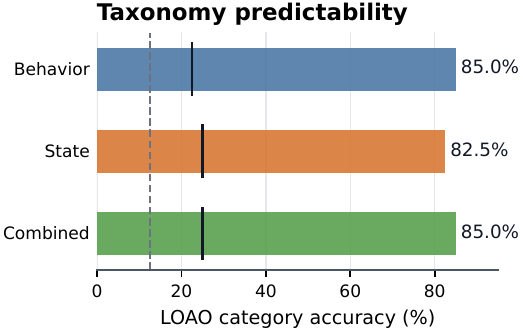}
  \caption{Held-out mechanism-category recovery. Ticks mark 95th percentiles over 5,000 balanced permutations.}
  \label{fig:taxonomy-predictability}
\end{figure}

\paragraph{Category recovery and evidentiary boundaries.}
The held-out evaluation in Figure~\ref{fig:taxonomy-predictability} provides the core empirical validation for \textbf{RQ1}: internal representation profiles alone recover the threat category of unseen attacks with 82.5\% accuracy, significantly exceeding the 95th-percentile permutation baseline of 25.0\% ($p < 0.001$). Behavioral outcomes alone achieve 85.0\% accuracy, and combining behavioral and state features maintains 85.0\%. This demonstrates that intermediate hidden states independently encode category-level structural signatures, without claiming an empirical advantage over behavioral classification.

Category recovery succeeds because related mechanisms redistribute internal representations across depth in characteristic patterns specific to their intervention surfaces (e.g., direct token optimization vs. conversational context vs. parameter edits). Because complete holdouts eliminate shared prompt artifacts, this transfer reflects genuine geometric organization rather than generic anomaly magnitude. Crucially, this confirms discriminant validity rather than causal localization; observed peaks identify diagnostic trace locations rather than causal intervention targets~\cite{meng2022rome}. This target-calibrated, multi-depth geometry provides the foundational signal exploited by the single-trace runtime auditor in Section~\ref{sec:audit-method}.
\section{\audit Design}
\label{sec:audit-method}

While Section~\ref{sec:cross-category} demonstrated that multi-depth hidden-state trajectories encode transferable structural signatures, that discovery relied on offline aggregation and matched clean--attack pairs. In real-world deployments, an auditor must evaluate individual inference traces without access to clean baselines or attack metadata. \audit operationalizes these empirical insights into a target-calibrated, candidate-only auditing framework operating under strict zero-oracle constraints.

\subsection{Design Requirements}
\label{sec:design-requirements}

The empirical findings in Section~\ref{sec:within-category} impose four strict design requirements:
(i) \textit{Candidate-only inference}: The auditor must operate strictly on the single observed runtime execution and target-model identifier, without requiring paired clean executions.
(ii) \textit{Universal detector architecture}: A single unified detector form must evaluate all 40 mechanisms without relying on attack category hints or threat-specific routing.
(iii) \textit{Target-specific calibration}: While the feature extraction pipeline is shared across architectures, benign reference distributions and decision thresholds must be calibrated locally to accommodate model-specific residual scales.
(iv) \textit{False-alarm discipline}: The auditor must sustain an operational, sub-percent false-positive rate across challenging benign distributions, including encoded texts, long-context passages, quoted untrusted content, and benign security discussions.

\begin{figure}[!t]
  \centering
  \includegraphics[width=0.72\columnwidth]{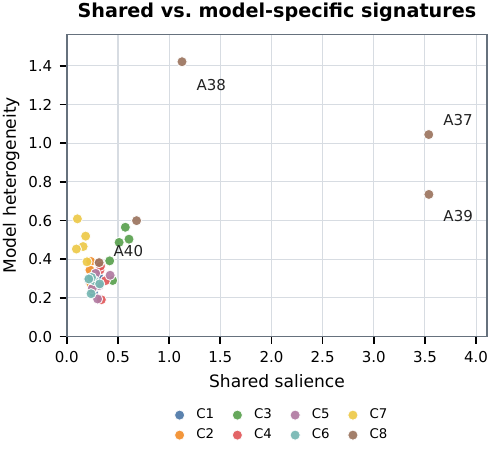}
  \caption{Shared layer-wise salience coexists with target-specific residual structure, motivating a common representation with target-specific calibration.}
  \label{fig:shared-vs-model-specific}
\end{figure}

These requirements dictate a principled division of architectural components (Figure~\ref{fig:shared-vs-model-specific}). The multi-depth feature extraction, random-projection basis, and hybrid two-gate decision logic are identical across all threat classes. In contrast, each target model maintains an independent benign reference bank, robust standardization scale, and operating threshold. This design reflects the fundamental representational finding of Section~\ref{sec:cross-category}: common depth-wise activation salience coexists with substantial target-specific baseline geometry.

\noindent\textbf{Measurement-to-design mapping.}
Every component of \audit directly addresses a measured empirical constraint: the absence of a universal peak layer necessitates multi-depth pooling; category-level transferability motivates retaining distributed trajectory geometry rather than scalar anomaly scores; persistent target residuals mandate target-specific normalization; and non-uniform threat profiles motivate a hybrid decision rule that couples supervised resemblance with unsupervised anomaly detection.

\subsection{State Representation}
\label{sec:state-representation}

Because the discovery analysis revealed that transferable signatures reside in relative depth trajectories rather than absolute layer indices or raw activation norms, \audit maps each execution onto a standardized relative-depth grid. For a candidate execution $x$ on target model $m$, the instrumentation pipeline extracts the residual stream activation vector at the final prompt token across all $L$ layers, matching the pre-generation boundary defined in Section~\ref{sec:within-representation}. To preserve coordinate geometry without overfitting to seen attack primitives, activations are compressed via a fixed Achlioptas random projection~\cite{achlioptas2003random}. Resampling these projected trajectories onto 21 normalized depth bins--fewer than the 25 layers of the shallowest target--standardizes the depth axis across architectures ($u \in [0, 1]$) without introducing interpolation artifacts.

The resulting representation combines the projected coordinate sequence with seven structural depth profiles: log activation norm, projected norm, coordinate mean magnitude, coordinate standard deviation, maximum coordinate magnitude, adjacent-layer cosine similarity, and adjacent-layer displacement. The coordinate trajectories retain localized representational shifts, while the structural profiles capture energy distribution, dispersion, and layer-to-layer transitions across depth.

To isolate adversarial displacement from model-specific activation scales, \audit standardizes each feature against a target-specific benign reference bank $\mathcal{B}_m$. Let $c_{mj}$ denote the median of feature $j$ over $\mathcal{B}_m$, and let $s_{mj}$ denote its inter-percentile range (10th--90th percentile), lower-bounded by $\epsilon > 0$ for numerical stability. Using medians and robust percentiles suppresses the influence of benign outliers without requiring attack data. The calibrated feature vector is computed as:
\begin{equation}
  z_j(x,m)=\operatorname{clip}\!\left(
    \frac{f_j(x)-c_{mj}}{s_{mj}},-20,20\right).
  \label{eq:target-calibrated-feature}
\end{equation}
Crucially, Equation~\eqref{eq:target-calibrated-feature} operates strictly on the candidate execution $x$ and the pre-computed benign reference $\mathcal{B}_m$; matched clean executions are required only during offline discovery and are entirely absent during deployment.

\begin{table*}[!t]
  \centering
  \small
  \setlength{\tabcolsep}{4.2pt}
  \renewcommand{\arraystretch}{1.08}
  \caption{Candidate-only auditing across 40 mechanisms and five targets. Panel A averages 40 LOAO folds per target; Panel B reports grouped holdouts with bootstrap 95\% half-widths. Boldface marks headline operating points.}
  \label{tab:reference-auditor}
  \begin{tabular}{@{}lrrrrr@{}}
    \toprule
    \multicolumn{6}{c}{Panel A: target-calibrated LOAO} \\
    \midrule
    Target model & AUROC & AUPRC & TPR (\%) & FPR (\%) & Max FPR (\%) \\
    \midrule
    Qwen3-4B & 0.994 & 0.944 & 76.7 & \textbf{0.46} & \textbf{1.54} \\
    Llama-3.2-3B & 0.990 & 0.925 & 69.3 & 0.57 & 1.91 \\
    DeepSeek-R1-7B & 0.989 & 0.898 & 80.2 & 1.55 & 5.23 \\
    Gemma-3-12B & \textbf{0.997} & 0.964 & 87.5 & 0.47 & 1.57 \\
    GPT-OSS-20B & \textbf{0.997} & \textbf{0.968} & \textbf{92.0} & 0.47 & 1.57 \\
    \midrule
    \multicolumn{6}{c}{Panel B: progressively harder grouped holdouts} \\
    \midrule
    Held-out unit & AUROC & AUPRC & TPR (\%) & FPR (\%) & Max FPR (\%) \\
    \midrule
    Attack (LOAO) & 0.993 $\pm$ 0.001 & 0.940 $\pm$ 0.006 & \textbf{81.1 $\pm$ 2.3} & \textbf{0.70 $\pm$ 0.07} & 2.37 $\pm$ 0.22 \\
    Category (LOCO) & 0.993 $\pm$ 0.002 & 0.986 $\pm$ 0.003 & \textbf{78.1 $\pm$ 5.3} & \textbf{0.66 $\pm$ 0.13} & 2.22 $\pm$ 0.45 \\
    Model + category & 0.915 $\pm$ 0.004 & 0.832 $\pm$ 0.010 & 5.0 $\pm$ 1.6 & 0.69 $\pm$ 0.12 & 2.32 $\pm$ 0.41 \\
    \bottomrule
  \end{tabular}
\end{table*}

\subsection{Hybrid Detection and Decision Policy}
\label{sec:hybrid-detector}

To provide comprehensive coverage across both familiar and novel threat primitives, \audit employs a hybrid two-gate detection architecture. The first gate deploys an ExtraTrees classifier~\cite{geurts2006extratrees} trained on multi-depth feature tables across observed attack mechanisms; its score $s_{\mathrm{ET}}(x)$ captures non-linear cross-depth geometric interactions indicative of known attack strategies. The second gate fits an unsupervised robust Mahalanobis distance estimator~\cite{lee2018mahalanobis} strictly on benign reference profiles; its score $s_{\mathrm{MD}}(x)$ quantifies departure from the target's nominal activation distribution. This dual structure ensures that an unfamiliar attack mechanism that diverges from the benign manifold can be detected even if it does not resemble previously observed training attacks.

Because the two gates evaluate complementary statistical evidence with disparate score distributions, a weighted linear blend could allow a strong anomaly in one view to be masked by the other. Instead, \audit issues an alert if either gate rejects the null hypothesis of benign execution. Under a total false-positive rate budget $\alpha$, each gate is allocated an independent budget of $\alpha/2$, with its decision threshold ($\tau_{\mathrm{ET}}, \tau_{\mathrm{MD}}$) set to the $(1 - \alpha/2)$ quantile over a validation split of challenging benign controls. The combined decision rule is:
\begin{equation}
  \widehat y(x)=
  \mathbf{1}\!\left[s_{\mathrm{ET}}(x)\geq\tau_{\mathrm{ET}}
  \;\lor\;s_{\mathrm{MD}}(x)\geq\tau_{\mathrm{MD}}\right].
  \label{eq:hybrid-audit-decision}
\end{equation}
This Bonferroni partition guarantees that the overall false-positive rate remains bounded by $\alpha$. For threshold-independent ranking and ROC-AUC evaluation, \audit outputs the maximum empirical benign percentile across the two gates.

\paragraph{Runtime inference and zero-oracle deployment.}
During offline training, permitted attacks train the ExtraTrees head, benign records fit the Mahalanobis gate, and a disjoint benign split calibrates target-specific thresholds. Attack identities and category labels define holdout partitions but are never provided as detector inputs. At runtime, \audit operates under strict zero-oracle constraints: evaluating a candidate execution requires solely the target-model identifier and the raw prompt activations, without access to clean counterparts, attack category labels, or downstream execution feedback. Generated alerts signify internal representation exposure relative to the calibrated benign baseline, adhering to the bounded evidentiary scope formalized in Section~\ref{sec:outcomes}. Implementation specifications, hyperparameter configurations, and split manifests appear in Appendix~\ref{app:audit-protocol}.
\section{Audit Evaluation}
\label{sec:audit-evaluation}

\begin{figure*}[!t]
  \centering
  \includegraphics[width=\textwidth]{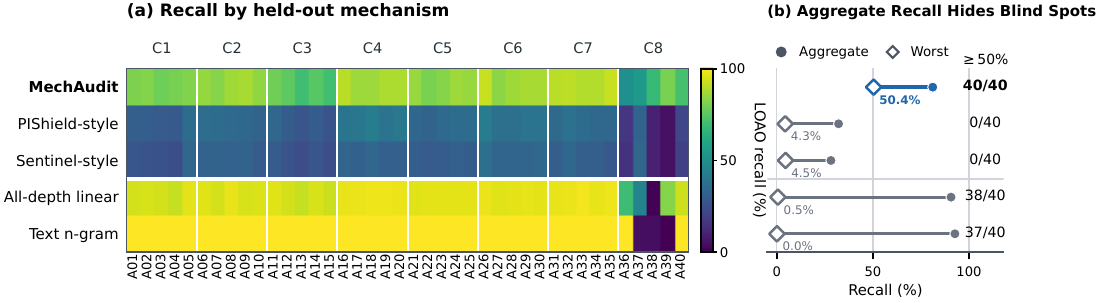}
\caption{Auditor comparison of mechanism-level recall under matched evaluation and a nominal 1\% family-wise false-positive budget. \textbf{(a)} Target-averaged LOAO recall. \textbf{(b)} Aggregate and worst-mechanism recall; right labels count mechanisms above 50\%. \audit alone maintains majority recall on all 40 mechanisms. Appendix Table~\ref{tab:detector-comparison-full} reports full metrics.}
  \label{fig:detector-coverage-comparison}
\end{figure*}

We evaluate \audit across 117,000 validation records to resolve \textbf{RQ2 (Generalization under Holdout)} and \textbf{RQ3 (Operational Boundaries)}. First, we measure single-trace auditing efficacy against unseen threat mechanisms and categories under strict zero-oracle constraints, comparing coverage against established baseline detectors. Second, we empirically delineate the evidentiary scope of an audit alert, establishing why representation-level exposure decouples from downstream task compromise.

\subsection{Unseen-Mechanism Auditing}
\label{sec:unseen-auditing}

\paragraph{Evaluation protocol.}
We evaluate 117,000 attack and benign-control records spanning five target architectures and all 40 mechanisms. Every evaluation enforces the candidate-only, zero-oracle setting: the feature extraction pipeline and hybrid decision rule are frozen across all mechanisms, while benign reference banks, robust scales, and decision thresholds are calibrated locally per target model. Leave-one-mechanism-out (LOMO/LOAO) removes the evaluated mechanism entirely from training; leave-one-category-out (LOCO) withholds all five mechanisms within its category. Mechanism and category labels govern holdout partitioning but are never exposed as detector inputs.

\paragraph{Unseen-mechanism generalization.}
Under LOMO evaluation, \audit achieves 0.993 AUROC and detects 81.1\% of unseen-attack executions at an operational false-positive rate (FPR) of 0.70\% (Table~\ref{tab:reference-auditor}). Under the more challenging LOCO protocol, AUROC remains 0.993, with \audit sustaining 78.1\% recall at 0.66\% FPR despite withholding the evaluated mechanism alongside all four category neighbors. This narrow 3.0-point recall gap confirms that \audit exploits a generalizable exposure signature that transcends category boundaries rather than overfitting to neighboring prompt templates.

This narrow LOMO--LOCO gap does not imply that threat categories are representatively indistinguishable. Rather, it reveals two complementary tiers of internal organization: fine-grained trajectory geometry that differentiates mechanism classes (as demonstrated in Section~\ref{sec:within-category}), and a broader attack-associated representational displacement that enables unified exposure auditing without category-specific supervision. Evaluating both holdout regimes is therefore critical: LOMO establishes out-of-mechanism transfer, while LOCO rules out local category-memorization shortcuts.

Performance remains consistent across calibrated target architectures, with AUROC ranging from 0.989 to 0.997, recall from 69.3\% to 92.0\%, and FPR from 0.46\% to 1.55\%, confirming that no single architecture skews aggregate results. Conversely, in a joint model-and-category holdout (transferring a source threshold to an uncalibrated target), AUROC remains high at 0.915, but recall drops to 5.0\% at the uncalibrated threshold. Thus, unseen-mechanism generalization transfers robustly across threat surfaces on calibrated targets, but strictly requires target-specific baseline calibration.

\paragraph{Mechanism-level coverage.}
While \audit provides unified detection across all 40 mechanisms without threat-specific routing, sensitivity exhibits expected variance across specific attack--target combinations. Appendix Figure~\ref{fig:projection-only-auditor-tpr-matrix} details all 200 LOMO cells, and Appendix Table~\ref{tab:reference-auditor-by-attack} provides per-mechanism recall averages.

\subsection{Auditor Comparison}
\label{sec:auditor-comparison}

\paragraph{Comparison protocol.}
All evaluated detectors are benchmarked on identical candidate records, grouped splits, benign controls, and target-calibration sets under a fixed 1\% family-wise false-positive budget. We implement two representative baselines adapted to internal representation monitoring: a single-layer linear classifier modeled after PIShield~\cite{zou2025pishield} and a complementary multi-layer probe modeled after Sentinel~\cite{liu2026sentinel}. In addition, we evaluate an all-depth linear baseline and a text $n$-gram classifier to benchmark flat activation separability and surface lexical regularities. Probe layer selection and fusion parameters are tuned on disjoint validation data, isolating architectural detection efficacy under a strictly controlled evaluation protocol.

\begin{figure*}[!t]
  \centering
  \includegraphics[width=0.85\textwidth]{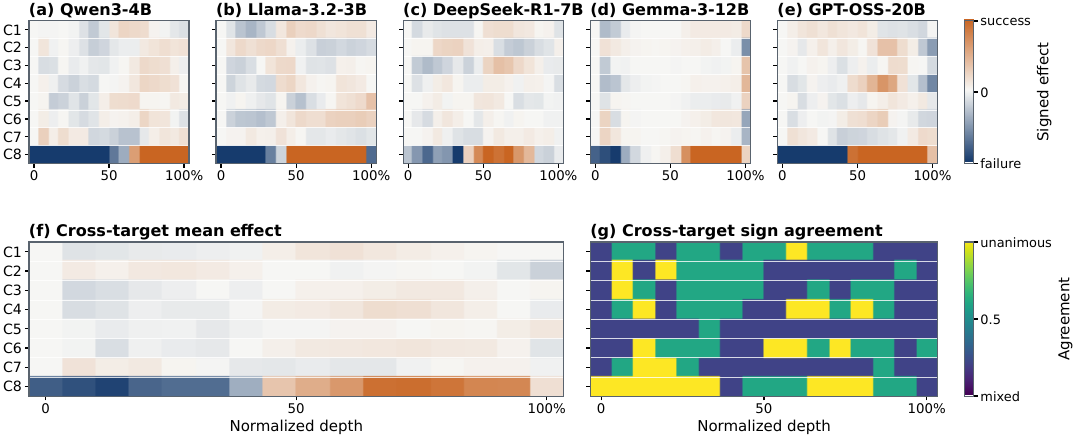}
  \caption{Successful-versus-failed state contrasts by target and in aggregate. Near-zero or inconsistent effects explain why exposure traces do not yield a reusable success marker.}
  \label{fig:audit-holdouts}
\end{figure*}

\begin{figure*}[!t]
  \centering
  \includegraphics[width=0.87\textwidth]{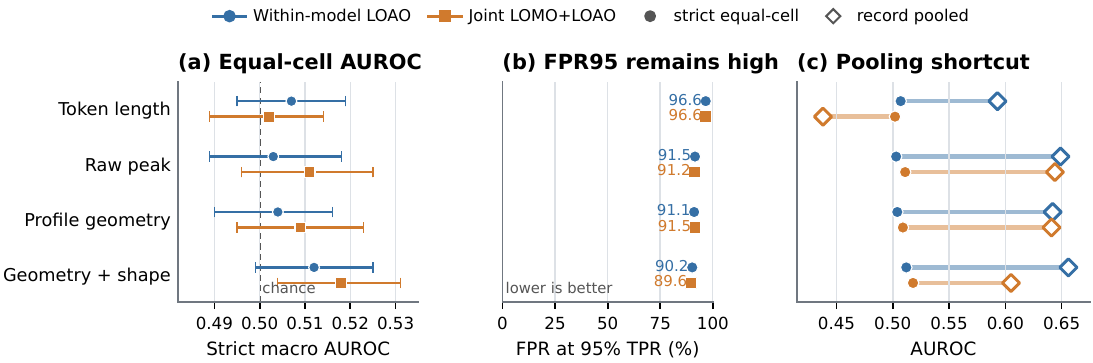}
  \caption{Protected-outcome claim boundary. \textbf{(a)} Equal-cell holdouts remain near chance. \textbf{(b)} Achieving 95\% TPR requires 89.6--96.6\% FPR. \textbf{(c)} Record pooling creates misleading separation. Appendix Table~\ref{tab:core-audit-holdout} reports complete values.}
  \label{fig:protected-outcome-claim-boundary}
\end{figure*}

\paragraph{Cross-mechanism transfer.}
Under LOMO evaluation, PIShield-style and Sentinel-style baselines achieve only 32.1\% and 28.1\% recall, respectively, whereas \audit achieves 81.1\% recall at a 0.70\% FPR. Under complete category holdouts (LOCO), this disparity persists: \audit retains 78.1\% recall, while baseline recall drops to 30.0\% and 26.0\% due to over-specialization to narrow threat signatures.

\paragraph{Coverage as a security property.}
In adversarial deployments, attackers strategically select the attack mechanism; high benchmark accuracy on trivial attacks provides no defense against a blind spot. Aggregate recall can dangerously mask vulnerabilities by rewarding detectors that over-index on dominant attack classes while leaving entire threat categories unprotected. To expose these vulnerabilities, we evaluate worst-mechanism recall and the number of mechanisms satisfying a 50\% recall floor. As illustrated in Figure~\ref{fig:detector-coverage-comparison}, all-depth linear and text $n$-gram detectors achieve high aggregate LOMO recall (90.7\% and 92.7\%) but collapse to 0.5\% and 0.0\% on their worst-case mechanisms, failing the 50\% threshold on multiple attack classes. In contrast, \audit maintains a target-averaged recall of at least 50.4\% across every evaluated mechanism, standing as the only detector to surpass the 50\% floor across all 40 mechanisms. Under category holdout, \audit's weakest category retains 63.8\% recall (versus 41.3\% for baselines). Thus, \audit's primary security advantage lies in eliminating complete mechanism blind spots rather than chasing aggregate score inflation.

\paragraph{Ablation analysis.}
Ablation experiments confirm that multi-depth projected profiles drive baseline sensitivity, while the unsupervised Mahalanobis gate provides critical tail coverage (Appendix Table~\ref{tab:candidate-auditor-ablation}). Replacing random-projected profiles with scalar hand-crafted features degrades ExtraTrees recall from 78.4\% to 56.9\%; coupling ExtraTrees with the Mahalanobis gate elevates recall from 78.4\% to 81.1\%. An oracle routing baseline that receives ground-truth intervention-surface labels achieves only 84.4\% recall, demonstrating that specialized attack metadata offers marginal benefit over our unified architecture. Furthermore, cryptographic audit checks across all 310 splits confirm zero train--test overlap in prompt hashes, base IDs, or feature vectors, ruling out data contamination shortcuts.

\subsection{Claim Boundary}
\label{sec:claim-boundary}

\paragraph{Exposure versus outcome compromise.}
An adversarial input induces internal representation shifts during prompt ingestion regardless of whether the model complies, generates a benign refusal, or suppresses an unauthorized tool call. To evaluate whether internal exposure traces can predict downstream attack success, we isolate 40,640 cryptographically joined behavior--state pairs across 194 attack--model cells under grouped evaluation~\cite{gulrajani2021domainbed,koh2021wilds}, ensuring that task difficulty and attack prevalence cannot act as confounding shortcuts.

\paragraph{Protected-outcome evaluation.}
Empirical state contrasts between successful and resisted attacks are sparse, inconsistent, and invert direction across architectures (Figure~\ref{fig:audit-holdouts}). Under equal-cell evaluation, internal geometry yields a near-chance macro AUROC of 0.512 within-model and 0.518 under joint model-plus-attack holdouts (Figure~\ref{fig:protected-outcome-claim-boundary}). Achieving a 95\% true-positive rate for success prediction requires an unacceptable false-positive rate of 89.6\%. In stark contrast to \audit's 0.993 exposure AUROC, internal representations fail to provide a generalizable success marker. This failure stems from a fundamental structural ordering: representational exposure occurs upstream during context processing, whereas attack success is determined downstream at the protected decision boundary. Naively pooling records produces an artificially inflated AUROC of 0.656 merely by learning cross-cell base rates, underscoring the necessity of equal-cell grouped evaluation.

\paragraph{Operational consequences.}
These findings establish an operational boundary for LLM security auditing: an \audit alert constitutes empirical evidence of internal adversarial exposure, but must not be treated as an autonomous verdict of system compromise. In production pipelines, alerts should trigger containment protocols, trace logging, or secondary semantic outcome validators rather than automated termination. Crucially, this limitation is not a defect of zero-oracle candidate deployment: even when an outcome probe is granted paired clean baselines, success prediction remains near chance under grouped evaluation. Bounding the evidentiary claim of auditing alerts prevents operational over-reliance while preserving reliable, early-stage threat detection.
\section{Related Work}
\label{sec:related-work}

\paragraph{Attack coverage and benchmarks.}
LLM security research spans diverse attack primitives, including input token optimization and iterative feedback search~\cite{zou2023gcg,sadasivan2024beast,chao2023pair,liu2024autodan}, semantic restructuring and carrier transformations~\cite{ding2024renellm,li2023deepinception,yuan2024cipherchat,jiang2024artprompt}, multi-turn context and persistent state manipulation~\cite{russinovich2025crescendo,greshake2023indirect,debenedetti2024agentdojo,zou2025poisonedrag,chen2024agentpoison}, and model-parameter or training backdoors that tamper with instruction data, parameter adapters, or internal reasoning steps~\cite{yan2024vpi,xu2024instructions,li2024badedit,wang2024badagent,guo2025darkmind}. Standardized benchmarks independently evaluate subsets of these threat surfaces: HarmBench and JailbreakBench evaluate safety policy violations under direct prompts~\cite{mazeika2024harmbench,chao2024jailbreakbench}; InjecAgent and TensorTrust measure instruction-integrity compromise in agentic workflows~\cite{zhan2024injecagent,toyer2023tensortrust}; and BackdoorLLM measures conditional trigger execution~\cite{li2025backdoorllm}. However, their underlying security invariants are fundamentally non-interchangeable: conflating safety refusal bypass with unauthorized tool invocation or poisoned retrieval corrupts evaluation denominators. Furthermore, existing benchmarks evaluate external behavioral endpoints without testing whether disparate attack surfaces exhibit transferable representation-level signatures. \study bridges these fragmented settings by formalizing a balanced 40-mechanism registry while enforcing threat-specific ground-truth outcomes and rigorous denominator hygiene.

\paragraph{White-box auditing and generalization.}
Representation-based defenses monitor intermediate hidden states, attention circuits, or reasoning trajectories to identify adversarial interactions~\cite{kirch2025features,zhang2025jbshield,he2026jailbreakscope,wen2025instructdetector,liu2026trajguard,guo2026traceguard}. PIShield inspects an isolated intermediate layer for prompt-injection detection~\cite{zou2025pishield}, while Sentinel combines complementary layer activations to detect direct jailbreaks~\cite{liu2026sentinel}. While these approaches demonstrate that internal representations are predictive within narrowly scoped threat models, they do not establish whether an identical detector architecture can generalize when an evaluated attack--or its entire functional category--is withheld during training. Standard random train--test splits risk rewarding localized lexical shortcuts or architecture-specific activation scales rather than generalizable representation geometry. In contrast, \study evaluates complete mechanism- and category-level holdouts under strict zero-oracle candidate deployment, testing whether a unified auditor can detect unseen threats spanning direct optimization, context manipulation, state poisoning, and training-time backdoors.

\paragraph{Evaluation under distribution shift and adaptive threats.}
Out-of-distribution generalization claims critically depend on the rigor of evaluation partitioning. Empirical studies in distribution shift demonstrate that random sampling significantly overstates model robustness when operational domains shift~\cite{gulrajani2021domainbed,koh2021wilds}, while representation-based safety filters frequently succumb to spurious lexical correlations or representation obfuscation~\cite{wang2026false,bailey2024obfuscated}. To eliminate these evaluation shortcuts, our methodology combines grouped task-and-model partitions, cell-level macro-averaging, cryptographic hash audits, and hard-negative calibration. Finally, we explicitly distinguish out-of-distribution transfer from adaptive robustness: adversaries equipped with detector gradient feedback or white-box oracle access can craft evasive perturbations to bypass activation monitors~\cite{nasr2026attacker}. In accordance with standard baseline threat models, we rigorously evaluate cross-mechanism generalization against non-adaptive adversaries, establishing the baseline boundaries of representation auditing before introducing adaptive game-theoretic dynamics.
\section{Conclusion}
\label{sec:conclusion}

Across 40 diverse mechanisms and five target architectures, \study demonstrates that heterogeneous LLM attacks leave transferable, structural signatures in internal representations. Rather than localized layer anomalies, this signal resides in target-calibrated geometric trajectories across model depth. Rigorous mechanism- and category-level holdouts confirm that these patterns reflect genuine threat structure rather than dataset or architectural scale artifacts.

Under candidate-only, zero-oracle deployment, \audit detects over 81\% of held-out attack executions at a 0.70\% false-positive rate, preserving sensitivity even across unseen threat categories. While linear and lexical baselines overfit to dominant attacks and suffer complete blind spots on out-of-distribution threats, \audit uniquely maintains target-averaged recall above 50\% across all 40 mechanisms--confirming that sustaining a coverage floor across the threat registry is the decisive security property.

For defense lifecycle maintenance, emerging threat mechanisms must first undergo complete holdout evaluation before entering training sets, preventing out-of-distribution auditing from silently degrading into in-distribution interpolation. Importantly, an \audit alert denotes internal representation-level exposure, delineating early detection from downstream task compromise or checkpoint tampering. Shifting white-box evaluation from average accuracy on familiar jailbreaks to worst-case holdout coverage provides a principled foundation for auditing evolving LLM threat surfaces without deploying specialized detectors for each individual exploit.

\nobalance

\section*{Ethical Considerations}

\paragraph{Stakeholders and risks.}
Evaluating 40 attack mechanisms in one framework benefits auditors and model developers, but it also concentrates techniques that were previously distributed across papers and repositories. The principal risks are easier reproduction of harmful attacks, exposure of annotators to harmful model outputs, and unintended effects on external systems or users. The study does not target production services, third-party accounts, or real users. Agent and retrieval experiments operate on disposable, access-controlled resources, and the corpus contains research identifiers rather than personal data, credentials, or user conversations.

\paragraph{Research safeguards.}
We minimize attack execution to the measurements needed for the stated research questions and separate behavioral judgment from raw-output access. Automated judgments reduce, but do not eliminate, annotation risk or error. Manual validation uses content warnings, the minimum context needed for adjudication, opt-out, and disagreement resolution. The experiments preserve refusals, failed constructors, inapplicable cases, and infrastructure failures rather than retrying until success.

\paragraph{Release and downstream use.}
The public artifact prioritizes source identifiers, frozen revisions, schemas, checksums, derived profiles, non-sensitive aggregates, and benign structural examples. Raw harmful requests and outputs, working poisoned records, and modified checkpoints remain in a controlled partition. This division supports verification of the paper's measurements without releasing a turnkey attack collection. Users remain responsible for authorization, least privilege, sandboxing, and downstream review.

\section*{Open Science}

\paragraph{Review access.}
The anonymous review package is available at \url{https://anonymous.4open.science/r/artifact-review-64a85ea5-F79E/}. It contains the frozen bundle, checksums, and claim-to-artifact index without identifying authors or collecting reviewer information.

\paragraph{Artifact contents.}
The package contains the source registry, taxonomy and adjudication log, model identifiers, grouped splits, execution manifests, trace schemas, out-of-fold predictions, aggregate tables, and figure and table scripts. Run manifests record status, configurations, hashes, job identifiers, outputs, and headline metrics. Exact input hashes connect behavioral records to white-box traces for auditing joins and denominators.

\paragraph{Reproduction and restrictions.}
Setup instructions rebuild holdout tests, tables, and figures from frozen intermediate records. Public materials contain non-sensitive metadata, derived measurements, benign examples, and deterministic analysis code; controlled components follow the safeguards in Ethical Considerations. After acceptance, the verified artifact will be published at a stable URL.

\bibliographystyle{plain} 
\bibliography{references}

\appendix
\onecolumn
\section{Frozen 40-Mechanism Attack Registry}
\label{app:registry}

The frozen registry fixes each documented mechanism, primary category, attack surface, construction procedure, and duration before any internal trace is inspected. Its human-readable form appears in Table~\ref{tab:full-registry}, while the accompanying artifact provides the machine-readable registry. The source field identifies the bibliography entry used to verify the method description. Execution fidelity, artifact revision, alternates, and exclusions are recorded in the machine-readable evidence package through the source-artifact gate described in Appendix~\ref{app:measurement-protocol}. A failed gate may activate only an alternate from the same category that was fixed before the reported analyses; it cannot change a category definition or reduce another category's allocation.

\setlength{\LTcapwidth}{\textwidth}
\small
\setlength{\tabcolsep}{3.5pt}
\begin{longtable}{@{}C{0.045\textwidth}C{0.045\textwidth}L{0.18\textwidth}L{0.18\textwidth}L{0.29\textwidth}C{0.11\textwidth}@{}}
  \caption{Frozen 40-mechanism attack registry. The registry balances
  eight primary categories with five attack mechanisms each. Surface records
  where the attack acts, construction records how it is created, and duration
  distinguishes request, session, cross-session, and deployment scope.}
  \label{tab:full-registry}\\
  \toprule
  ID & Cat. & Attack & Surface
  & Construction & Duration \\
  \midrule
  \endfirsthead
  \multicolumn{6}{l}{\small\itshape Table~\ref{tab:full-registry} continued.}\\
  \toprule
  ID & Cat. & Attack & Surface
  & Construction & Duration \\
  \midrule
  \endhead
  \midrule
  \multicolumn{6}{r}{\small\itshape Continued on next page.}\\
  \endfoot
  \bottomrule
  \endlastfoot
  A01 & \Cat{1} & GCG & current input & gradients and internal objective & request \\
  A02 & \Cat{1} & BEAST & current input & next-token objective and search & request \\
  A03 & \Cat{1} & COLD-Attack & current input & gradients and logits & request \\
  A04 & \Cat{1} & ASETF & current input & surrogate embeddings and gradients & request \\
  A05 & \Cat{1} & AdvPrompter & current input & learned surrogate objective & request \\
  \addlinespace
  A06 & \Cat{2} & PAIR & current input & target responses and judge feedback & request \\
  A07 & \Cat{2} & TAP & current input & target feedback with tree pruning & request \\
  A08 & \Cat{2} & AutoDAN & current input & target fitness and candidate evolution & request \\
  A09 & \Cat{2} & Rainbow Teaming & current input & behavioral feedback and diversity objective & request \\
  A10 & \Cat{2} & ECLIPSE & current input & target responses and iterative search & request \\
  \addlinespace
  A11 & \Cat{3} & ReNeLLM & current input & semantic rewrite and scenario nesting & request \\
  A12 & \Cat{3} & DeepInception & current input & nested narrative framing & request \\
  A13 & \Cat{3} & PAP & current input & persuasive strategy & request \\
  A14 & \Cat{3} & DrAttack & current input & semantic decomposition and reconstruction & request \\
  A15 & \Cat{3} & DRA & current input & disguised reconstruction & request \\
  \addlinespace
  A16 & \Cat{4} & CipherChat & current input & cipher carrier & request \\
  A17 & \Cat{4} & ArtPrompt & current input & typographic and ASCII-art carrier & request \\
  A18 & \Cat{4} & CodeChameleon & current input & code-structured carrier & request \\
  A19 & \Cat{4} & M2S & current input & structured single-request serialization & request \\
  A20 & \Cat{4} & FlipAttack & current input & order transformation & request \\
  \addlinespace
  A21 & \Cat{5} & Crescendo & dialogue history & progressive escalation & session \\
  A22 & \Cat{5} & Foot-In-The-Door & dialogue history & incremental commitment & session \\
  A23 & \Cat{5} & DAMON & dialogue history & adaptive multi-turn manipulation & session \\
  A24 & \Cat{5} & RACE & dialogue history & refusal recovery and escalation & session \\
  A25 & \Cat{5} & PE-CoA & dialogue history & pattern-enhanced multi-turn construction & session \\
  \addlinespace
  A26 & \Cat{6} & Indirect Prompt Injection & external runtime content & untrusted-content instruction & request / session \\
  A27 & \Cat{6} & HouYi & external runtime content & application-mediated instruction & request / session \\
  A28 & \Cat{6} & Adaptive IPI & external runtime content & adaptive indirect instruction & request / session \\
  A29 & \Cat{6} & TopicAttack & external runtime content & topic-conditioned indirect instruction & request / session \\
  A30 & \Cat{6} & AgentVigil & tool or agent channel & agent-channel instruction & request / session \\
  \addlinespace
  A31 & \Cat{7} & PoisonedRAG & external state & retrieval-corpus poisoning & cross-session \\
  A32 & \Cat{7} & AgentPoison & external state & agent knowledge or memory poisoning & cross-session \\
  A33 & \Cat{7} & MINJA & external state & long-term memory injection & cross-session \\
  A34 & \Cat{7} & Phantom & external state & retrieval-corpus poisoning & cross-session \\
  A35 & \Cat{7} & Machine Against the RAG & external state & adversarial retrieval records & cross-session \\
  \addlinespace
  A36 & \Cat{8} & Virtual Prompt Injection & model state & poisoned training or parameter state & deployment \\
  A37 & \Cat{8} & Universal Jailbreak Backdoors & model state & trigger-conditioned training & deployment \\
  A38 & \Cat{8} & Instructions as Backdoors & model state & instruction-triggered training & deployment \\
  A39 & \Cat{8} & BadEdit & model state & parameter editing & deployment \\
  A40 & \Cat{8} & BadAgent & model state & backdoored agent training & deployment \\
\end{longtable}
\normalsize

\subsection{Admission and Alternate Policy}

A record is eligible for the final measurement matrix only if four conditions hold: the primary source describes an independent attack method; the method can be mapped by the frozen decision rule; the protected execution can be expressed under one of the security tracks in Section~\ref{sec:outcomes}; and the essential method can be instantiated at fidelity F1 or higher. Benchmark suites, environments, datasets, broad attack collections, and prompt variants do not occupy mechanism slots. The source-artifact inspection must be completed without consulting attack success or internal measurements from this study, and its result must accompany any fidelity-dependent claim.

The protocol permits only a frozen alternate list established before the reported analyses. Each entry must contain a source URL, proposed primary and secondary labels, and the exact admission condition that would activate it. An alternate can replace only a record that fails for a documented scientific or technical reason. Failure on a particular model does not automatically remove a mechanism: the corresponding cell is marked unsupported, and category summaries use the common eligible matrix together with a missingness sensitivity analysis. All exclusions, substitutions, and unsupported cells must remain in the released registry.

\subsection{Annotation Audit Trail}

For each mechanism, the coding form records the attack surface, construction procedure, duration, required access, and protected security property. Annotators quote the source requirement supporting each field and record a plausible alternative category when applicable. The annotation interface hides all traces and outcomes produced by this study. The audit trail preserves both independent labels, a disagreement code, the adjudicated decision, and a short rationale. Category-level analysis begins only after this log and the measurement protocol are frozen.

\section{Behavioral Outcome Atlases}
\label{app:behavioral-atlas-values}

\begin{figure}[!ht]
  \centering
  \includegraphics[width=0.84\textwidth]{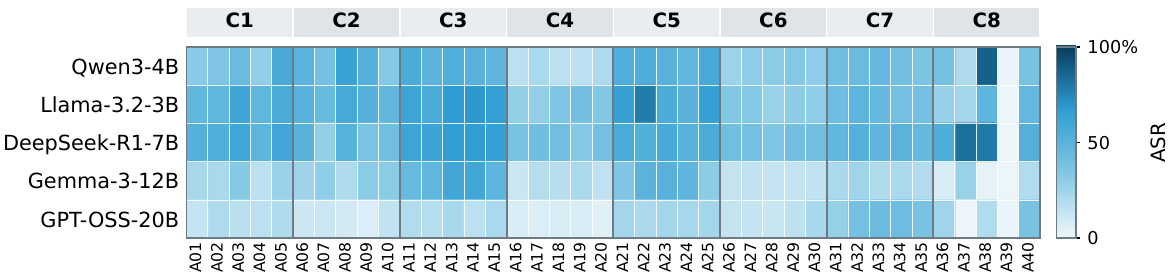}
  \caption{ASR across 40 attack mechanisms and five target models. Rows follow the frozen C1--C8 taxonomy; darker cells indicate higher applicable-response ASR.}
  \label{fig:behavioral-outcome-atlas}
\end{figure}

\begin{figure}[!ht]
  \centering
  \includegraphics[width=0.84\textwidth]{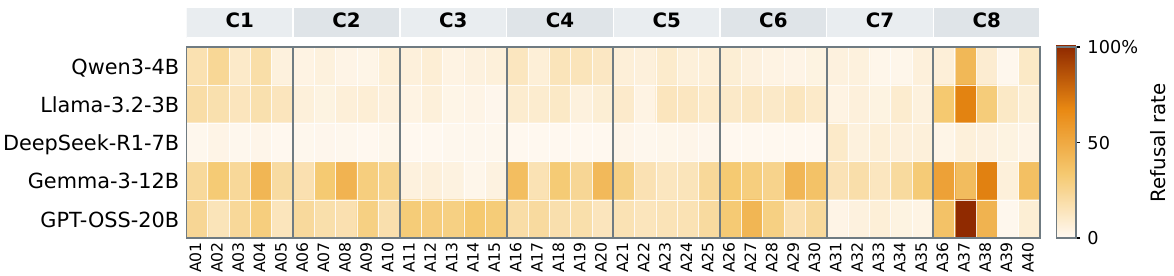}
  \caption{Refusal rates across 40 mechanisms and five targets; darker cells indicate more refusals.}
  \label{fig:behavioral-refusal-atlas-appendix}
\end{figure}

The complete numerical matrix appears in Table~\ref{tab:behavioral-atlas-values}. Applicable-response ASR and refusal over all judged responses appear in Figures~\ref{fig:behavioral-outcome-atlas} and~\ref{fig:behavioral-refusal-atlas-appendix}, respectively.

\section{Target-Specific Internal-State Atlases}
\label{app:target-model-depth-atlases}

\begin{figure}[!ht]
  \centering
  \includegraphics[width=0.88\textwidth]{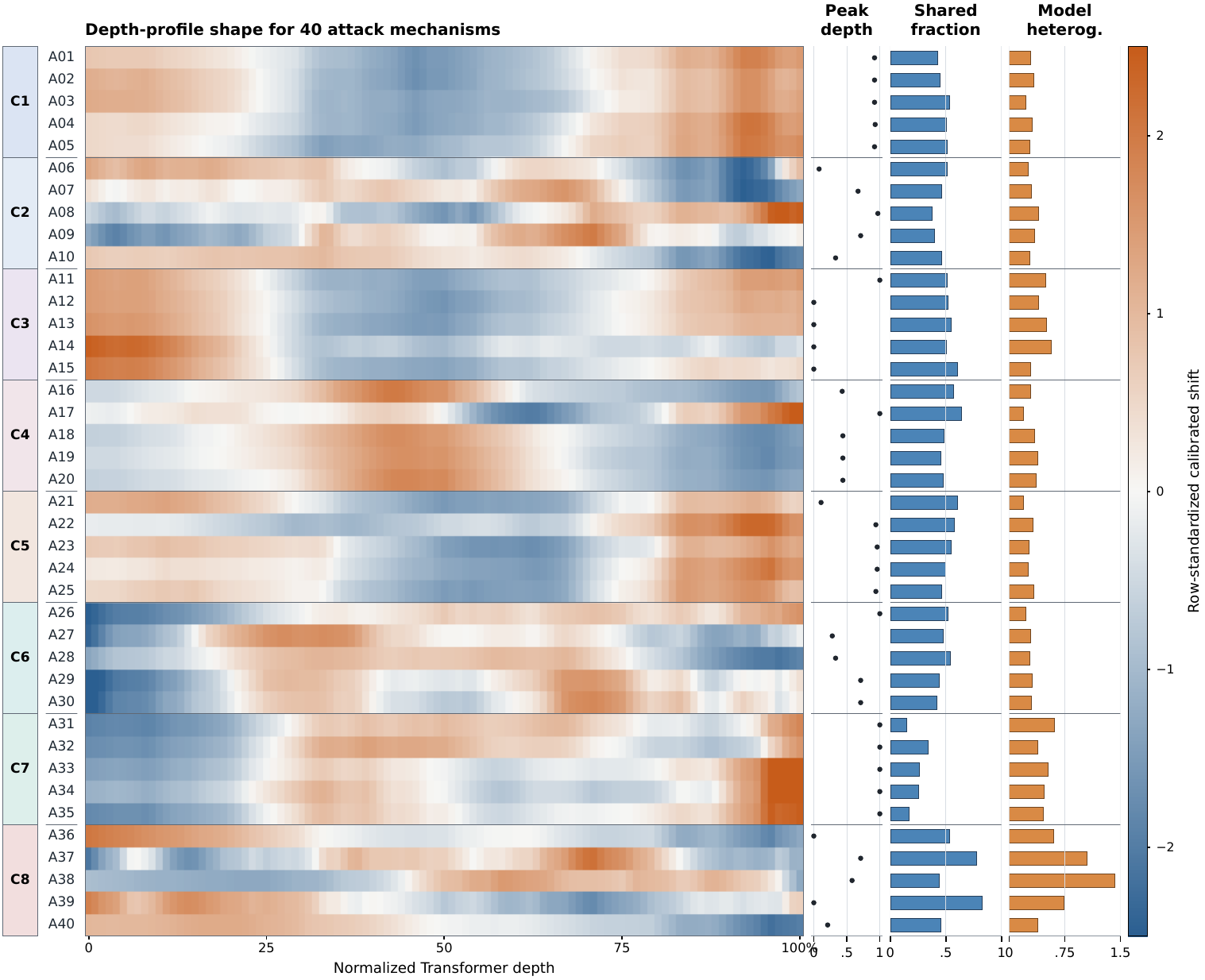}
  \caption{Mechanism-level depth atlas for 40 attacks (100,000 matched clean--attack pairs). Rows follow C1--C8; colors show calibrated profile shape over normalized depth, and side summaries report peak depth, shared fraction, and model heterogeneity.}
  \label{fig:mechanism-depth-atlas}
\end{figure}

Figure~\ref{fig:mechanism-depth-atlas} retains all 40 mechanisms on a common normalized-depth axis. Its side panels summarize peak depth, shared fraction, and model heterogeneity, exposing which signatures persist across targets without repeating a full-page atlas for every architecture.

\begin{figure}[!ht]
  \centering
  \includegraphics[width=0.88\textwidth]{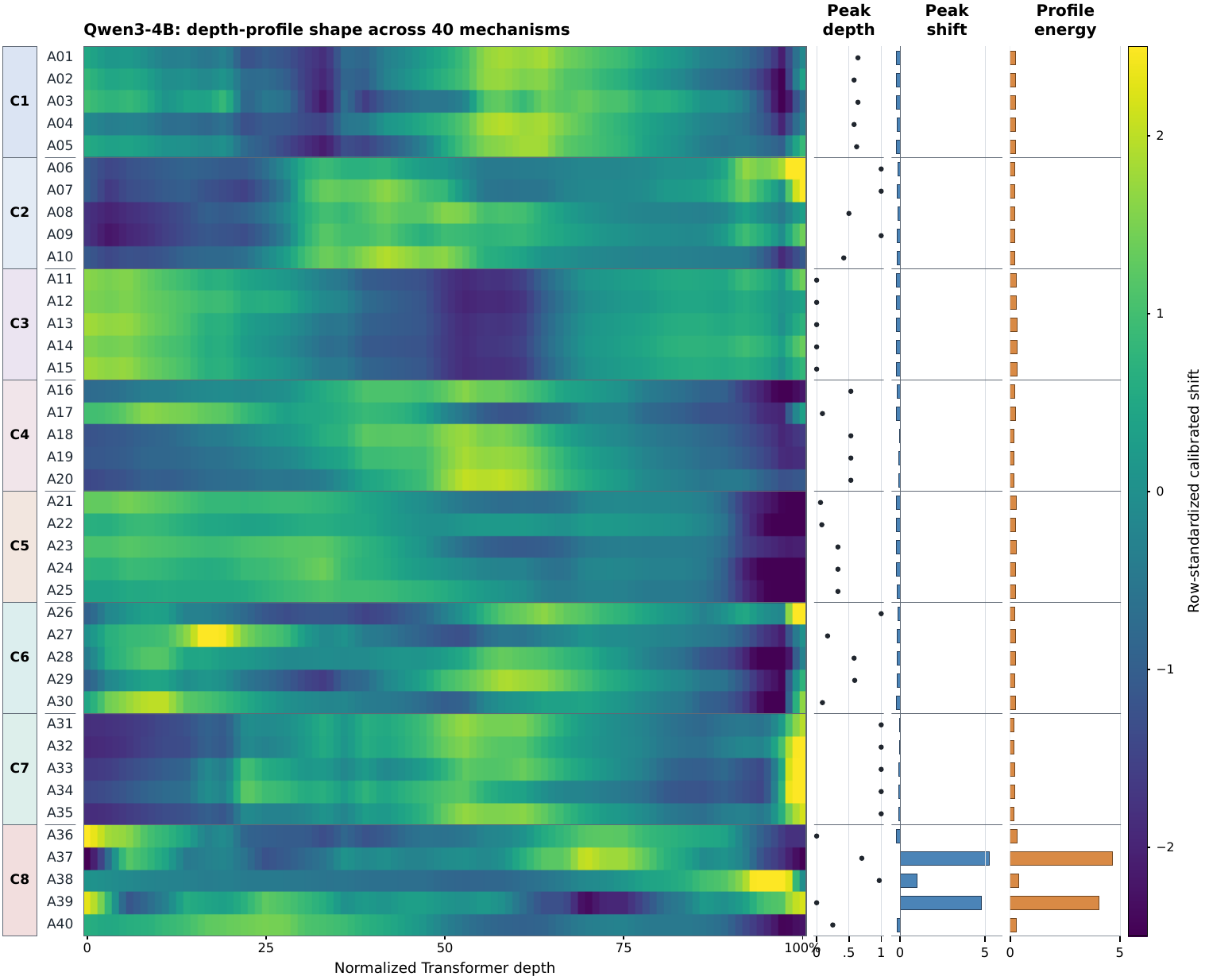}
  \caption{Target-specific internal-state depth atlas for Qwen3-4B. Rows A01--A40 follow C1--C8. The heatmap reports the row-standardized within-model calibrated cosine-displacement profile; the side panels report peak depth, peak calibrated shift, and profile energy. Even with architecture fixed, peaks span model depth and profile shapes differ within categories, motivating the full calibrated profile rather than a single peak layer or magnitude.}
  \label{fig:target-depth-atlas-qwen}
\end{figure}

Figure~\ref{fig:target-depth-atlas-qwen} gives the corresponding single-target view for Qwen3-4B.

The mechanism-level profiles vary in shape, peak depth, shared fraction, and model heterogeneity. This decomposition supports the paper's audit boundary: normalized profile geometry is useful for mechanism-level triage, but a raw layer index or magnitude threshold cannot be transferred unchanged across architectures.

\twocolumn
\section{Extended Cross-Model and Audit Diagnostics}
\label{app:extended-diagnostics}

\begin{table*}[!t]
  \centering
  \small
  \setlength{\tabcolsep}{1.2pt}
  \renewcommand{\arraystretch}{0.8}
  \caption{Unified candidate-only auditor by attack under within-model LOAO evaluation. Each target cell reports TPR at its calibrated threshold. The detector receives no attack label, intervention-surface label, protected-outcome label, or matched-clean trace at test time.}
  \label{tab:reference-auditor-by-attack}
  \begin{tabular*}{\textwidth}{@{\extracolsep{\fill}}l@{\hspace{2pt}}lrrrrrrrrr@{}}
    \toprule
    Attack & Cat. & \multicolumn{5}{c}{TPR by target (\%)} & \multicolumn{4}{c}{Cross-target summary} \\
    \cmidrule(lr){3-7}\cmidrule(l){8-11}
    & & Qwen & Llama & DeepSeek & Gemma & GPT-OSS & Mean & Worst & AUROC & Max FPR \\
    \midrule
    A01 GCG & C1 & 80.4 & 58.0 & 80.4 & 83.9 & 99.1 & 80.4 & 58.0 & 0.993 & 4.88 \\
    A02 BEAST & C1 & 82.1 & 58.9 & 89.3 & 77.7 & 99.1 & 81.4 & 58.9 & 0.994 & 5.85 \\
    A03 COLD-Attack & C1 & 73.2 & 64.3 & 71.4 & 81.2 & 99.1 & 77.9 & 64.3 & 0.993 & 4.88 \\
    A04 ASETF & C1 & 82.1 & 62.5 & 81.2 & 73.2 & 99.1 & 79.6 & 62.5 & 0.994 & 4.88 \\
    A05 AdvPrompter & C1 & 78.6 & 48.2 & 84.8 & 93.8 & 99.1 & 80.9 & 48.2 & 0.994 & 4.39 \\
    \addlinespace[1.5pt]
    A06 PAIR & C2 & 85.7 & 70.5 & 77.7 & 88.4 & 98.2 & 84.1 & 70.5 & 0.994 & 5.37 \\
    A07 TAP & C2 & 77.7 & 70.5 & 75.0 & 87.5 & 98.2 & 81.8 & 70.5 & 0.994 & 5.37 \\
    A08 AutoDAN & C2 & 86.6 & 65.2 & 82.1 & 94.6 & 100.0 & 85.7 & 65.2 & 0.995 & 5.37 \\
    A09 Rainbow Teaming & C2 & 89.3 & 81.2 & 80.4 & 84.8 & 98.2 & 86.8 & 80.4 & 0.994 & 5.37 \\
    A10 ECLIPSE & C2 & 80.4 & 83.0 & 77.7 & 86.6 & 90.2 & 83.6 & 77.7 & 0.994 & 4.88 \\
    \addlinespace[1.5pt]
    A11 ReNeLLM & C3 & 69.6 & 54.5 & 78.6 & 89.3 & 100.0 & 78.4 & 54.5 & 0.991 & 5.37 \\
    A12 DeepInception & C3 & 72.3 & 31.2 & 82.1 & 91.1 & 99.1 & 75.2 & 31.2 & 0.991 & 5.85 \\
    A13 PAP & C3 & 51.8 & 51.8 & 67.0 & 75.9 & 100.0 & 69.3 & 51.8 & 0.990 & 5.37 \\
    A14 DrAttack & C3 & 66.1 & 43.8 & 67.9 & 93.8 & 98.2 & 73.9 & 43.8 & 0.992 & 4.39 \\
    A15 DRA & C3 & 56.2 & 58.9 & 62.5 & 72.3 & 99.1 & 69.8 & 56.2 & 0.990 & 4.88 \\
    \addlinespace[1.5pt]
    A16 CipherChat & C4 & 81.2 & 88.4 & 88.4 & 91.1 & 99.1 & 89.6 & 81.2 & 0.995 & 4.88 \\
    A17 ArtPrompt & C4 & 71.4 & 83.0 & 86.6 & 86.6 & 100.0 & 85.5 & 71.4 & 0.995 & 5.37 \\
    A18 CodeChameleon & C4 & 82.1 & 78.6 & 87.5 & 83.9 & 94.6 & 85.4 & 78.6 & 0.995 & 5.37 \\
    A19 M2S & C4 & 75.0 & 95.5 & 88.4 & 84.8 & 94.6 & 87.7 & 75.0 & 0.996 & 5.37 \\
    A20 FlipAttack & C4 & 76.8 & 86.6 & 87.5 & 92.9 & 95.5 & 87.9 & 76.8 & 0.996 & 5.37 \\
    \addlinespace[1.5pt]
    A21 Crescendo & C5 & 83.0 & 68.8 & 81.2 & 90.2 & 100.0 & 84.6 & 68.8 & 0.994 & 5.37 \\
    A22 Foot-In-The-Door & C5 & 76.8 & 69.6 & 73.2 & 79.5 & 100.0 & 79.8 & 69.6 & 0.994 & 4.88 \\
    A23 DAMON & C5 & 79.5 & 67.9 & 76.8 & 85.7 & 100.0 & 82.0 & 67.9 & 0.994 & 5.37 \\
    A24 RACE & C5 & 81.2 & 79.5 & 78.6 & 87.5 & 100.0 & 85.4 & 78.6 & 0.995 & 5.85 \\
    A25 PE-CoA & C5 & 84.8 & 83.0 & 79.5 & 75.0 & 100.0 & 84.5 & 75.0 & 0.994 & 5.37 \\
    \addlinespace[1.5pt]
    A26 Indirect Prompt Injection & C6 & 87.5 & 87.5 & 87.5 & 97.3 & 99.1 & 91.8 & 87.5 & 0.996 & 5.37 \\
    A27 HouYi & C6 & 83.0 & 65.2 & 78.6 & 86.6 & 99.1 & 82.5 & 65.2 & 0.994 & 5.37 \\
    A28 Adaptive IPI & C6 & 77.7 & 79.5 & 82.1 & 91.1 & 99.1 & 85.9 & 77.7 & 0.995 & 4.88 \\
    A29 TopicAttack & C6 & 75.9 & 83.0 & 89.3 & 88.4 & 98.2 & 87.0 & 75.9 & 0.995 & 5.85 \\
    A30 AgentVigil & C6 & 81.2 & 75.9 & 86.6 & 91.1 & 99.1 & 86.8 & 75.9 & 0.995 & 5.37 \\
    \addlinespace[1.5pt]
    A31 PoisonedRAG & C7 & 94.6 & 68.8 & 88.4 & 97.3 & 100.0 & 89.8 & 68.8 & 0.995 & 5.37 \\
    A32 AgentPoison & C7 & 90.2 & 82.1 & 80.4 & 98.2 & 100.0 & 90.2 & 80.4 & 0.996 & 4.88 \\
    A33 MINJA & C7 & 84.8 & 73.2 & 85.7 & 98.2 & 100.0 & 88.4 & 73.2 & 0.995 & 4.88 \\
    A34 Phantom & C7 & 84.8 & 82.1 & 75.9 & 100.0 & 100.0 & 88.6 & 75.9 & 0.995 & 4.88 \\
    A35 Machine Against the RAG & C7 & 90.2 & 85.7 & 85.7 & 96.4 & 100.0 & 91.6 & 85.7 & 0.995 & 5.37 \\
    \addlinespace[1.5pt]
    A36 Virtual Prompt Injection & C8 & 76.8 & 12.5 & 67.0 & 52.7 & 42.9 & 50.4 & 12.5 & 0.981 & 5.37 \\
    A37 Universal Jailbreak Backdoors & C8 & 71.4 & 4.5 & 92.0 & 97.3 & 4.5 & 53.9 & 4.5 & 0.991 & 5.37 \\
    A38 Instructions as Backdoors & C8 & 0.0 & 99.1 & 63.4 & 92.0 & 83.0 & 67.5 & 0.0 & 0.988 & 5.85 \\
    A39 BadEdit & C8 & 81.7 & 100.0 & 94.6 & 100.0 & 22.6 & 79.8 & 22.6 & 0.995 & 4.88 \\
    A40 BadAgent & C8 & 63.4 & 70.5 & 67.0 & 81.2 & 71.4 & 70.7 & 63.4 & 0.992 & 5.37 \\
    \bottomrule
  \end{tabular*}
\end{table*}

\begin{figure}[!t]
  \centering
  \includegraphics[width=0.74\columnwidth]{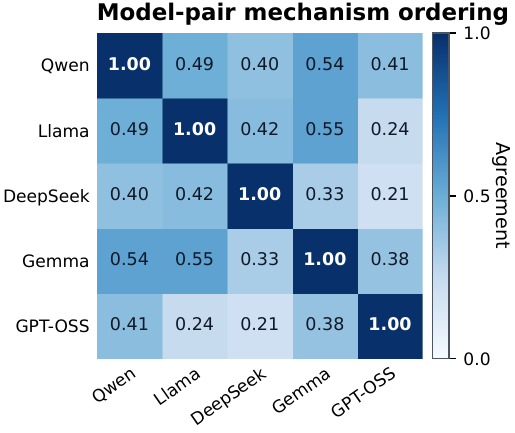}
  \caption{Cross-model agreement in calibrated mechanism ordering. Each cell averages feature-wise Spearman correlations across all 40 mechanisms.}
  \label{fig:model-pair-feature-ordering}
\end{figure}

\begin{table*}[!t]
  \centering
  \small
  \setlength{\tabcolsep}{3.5pt}
  \renewcommand{\arraystretch}{1.03}
  \caption{Complete numeric cross-model structure results. Panel A decomposes state variation; Panel B reports leave-one-attack-out taxonomy recovery; Panel C reports feature-wise mechanism-ordering agreement. Boldface marks the dominant component, best held-out accuracy, and strongest and weakest mean agreement.}
  \label{tab:full-crossmodel-structure}

  \textbf{Panel A: state-variation decomposition}\vspace{1pt}
  \begin{tabularx}{\textwidth}{@{}L{0.30\textwidth}L{0.23\textwidth}C{0.12\textwidth}C{0.12\textwidth}C{0.12\textwidth}@{}}
    \toprule
    Feature & Scale & Model (\%) & Attack (\%) & Inter. (\%) \\
    \midrule
    Cosine peak value & Raw & 6.5 & \textbf{86.5} & 6.9 \\
    Relative-$\ell_2$ peak value & Raw & 11.6 & \textbf{77.6} & 10.7 \\
    Cosine peak depth & Raw & \textbf{47.0} & 40.2 & 12.8 \\
    Cosine peak depth & Within-model calibrated & -- & \textbf{78.6} & 21.4 \\
    Relative-$\ell_2$ peak depth & Within-model calibrated & -- & \textbf{69.7} & 30.3 \\
    \bottomrule
  \end{tabularx}

  \vspace{0.45em}
  \textbf{Panel B: leave-one-attack-out taxonomy recovery}\vspace{1pt}
  \begin{tabularx}{\textwidth}{@{}L{0.28\textwidth}C{0.15\textwidth}C{0.15\textwidth}C{0.15\textwidth}C{0.12\textwidth}@{}}
    \toprule
    Evidence & Feat./model & LOAO acc. (\%) & Perm. q95 (\%) & $p$ \\
    \midrule
    Behavior only & 2 & \textbf{85.0} & 22.5 & $<.001$ \\
    Internal state only & 8 & 82.5 & 25.0 & $<.001$ \\
    Behavior + state & 10 & \textbf{85.0} & 25.0 & $<.001$ \\
    \bottomrule
  \end{tabularx}

  \vspace{0.45em}
  \textbf{Panel C: model-pair mechanism agreement}\vspace{1pt}
  \begin{tabularx}{\textwidth}{@{}L{0.24\textwidth}C{0.09\textwidth}C{0.10\textwidth}@{\hspace{0.02\textwidth}}L{0.24\textwidth}C{0.09\textwidth}C{0.10\textwidth}@{}}
    \toprule
    Model pair & Mean $\rho$ & Range &
    Model pair & Mean $\rho$ & Range \\
    \midrule
    Llama--Gemma & \textbf{0.55} & 0.27--0.80 & Qwen--DeepSeek & 0.40 & 0.04--0.64 \\
    Qwen--Gemma & 0.54 & 0.31--0.69 & Gemma--GPT-OSS & 0.38 & 0.13--0.83 \\
    Qwen--Llama & 0.49 & -0.06--0.83 & DeepSeek--Gemma & 0.33 & -0.04--0.73 \\
    Llama--DeepSeek & 0.42 & 0.05--0.63 & Llama--GPT-OSS & 0.24 & 0.02--0.54 \\
    Qwen--GPT-OSS & 0.41 & 0.07--0.64 & DeepSeek--GPT-OSS & \textbf{0.21} & -0.41--0.60 \\
    \bottomrule
  \end{tabularx}

  \vspace{2pt}
  \emph{Note.} Chance is 12.5\% for the eight-category recovery task; permutation summaries use 5,000 balanced label permutations.
\end{table*}

This appendix retains supporting figures whose quantitative conclusions are reported in the main-paper tables. All panels use the same frozen features, splits, and denominators as Section~\ref{sec:cross-category} and Section~\ref{sec:audit-evaluation}.

Figure~\ref{fig:model-pair-feature-ordering} makes the residual target dependence explicit: no off-diagonal model pair preserves uniformly high mechanism ordering, and agreement is weakest for pairs involving GPT-OSS.

\subsection{Cross-Target Signature Diagnostics}

\subsection{Detailed Cross-Target Interpretation}

For standardized feature vector $z_{a,m}\in\mathbb{R}^{10}$, the shared-salience and model-heterogeneity summaries are
\begin{equation}
 \begin{aligned}
 S_a &= \left\|\frac{1}{|\mathcal{M}|}\sum_m z_{a,m}\right\|_2/\sqrt{10},\\
 H_a &= \left(\frac{1}{10|\mathcal{M}|}\sum_m\left\|z_{a,m}-\bar z_a\right\|_2^2\right)^{1/2}.
 \end{aligned}
 \label{eq:shared-specific}
\end{equation}
No model pair preserves architecture-invariant ordering. Llama--Gemma and Qwen--Gemma have the largest mean feature correlations ($0.55$ and $0.54$), followed by Qwen--Llama ($0.49$) and Llama--DeepSeek ($0.42$). Agreement involving GPT-OSS is weaker: Llama--GPT-OSS is $0.24$ and DeepSeek--GPT-OSS is $0.21$.

A37 and A39 have the largest shared salience, but their abrupt carrier-induced shifts do not satisfy the length- and rare-token controls needed for a causal claim. A38 has the largest model-specific component, consistent with faithful conditional adapters on Qwen, Llama, and Gemma, weak trigger learning on DeepSeek, and global label collapse on GPT-OSS. C7 attacks tend to have low shared salience and comparatively high heterogeneity, consistent with model-specific retrieval and agent state.

The combined taxonomy representation misclassifies A36 as C1, A38 as C3, A40 as C7, and three non-C8 mechanisms as neighboring behavioral strata. These errors show that the taxonomy captures strong aggregate regularities without making mechanisms within a stratum execution-equivalent. The paired profiles use known targets and oracle clean counterparts, so taxonomy recovery remains a discriminant-validity test rather than a deployment detector.

\subsection{Complete Candidate-Only Auditor Results}

\begin{figure*}[!t]
  \centering
  \includegraphics[width=0.96\textwidth]{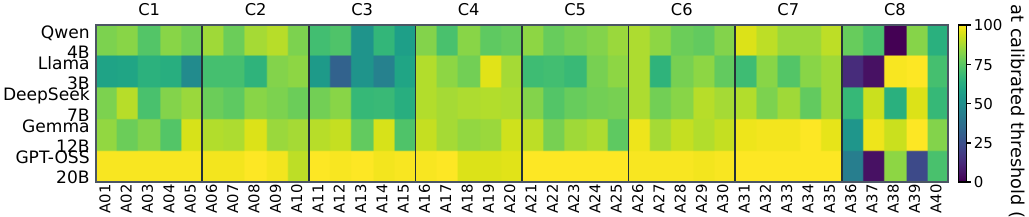}
  \caption{Unified candidate-only auditing recall across 40 attack mechanisms and five targets. Each cell reports TPR for an attack excluded from fitting; the same detector form is used for every mechanism with target-specific benign calibration.}
  \label{fig:projection-only-auditor-tpr-matrix}
\end{figure*}

\begin{figure}[!t]
  \centering
  \includegraphics[width=\columnwidth]{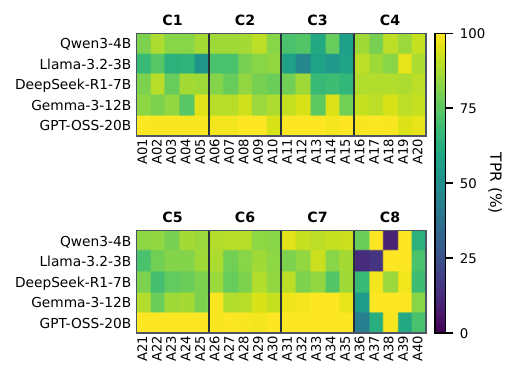}
  \caption{Oracle declared-surface sensitivity analysis across 40 mechanisms and five targets. This nondeployable diagnostic receives the intervention-surface label and an asset gate without benign parameter-edit controls.}
  \label{fig:surface-routed-auditor-matrix}
\end{figure}

The candidate-only hybrid applies one detector form to all 40 mechanisms, with cell-level and mechanism-level results in Figure~\ref{fig:projection-only-auditor-tpr-matrix} and Table~\ref{tab:reference-auditor-by-attack}. The separate sensitivity analysis in Figure~\ref{fig:surface-routed-auditor-matrix} receives the oracle intervention-surface label and is therefore not a deployment estimate.

\begin{figure*}[!t]
  \centering
  \includegraphics[width=0.84\textwidth]{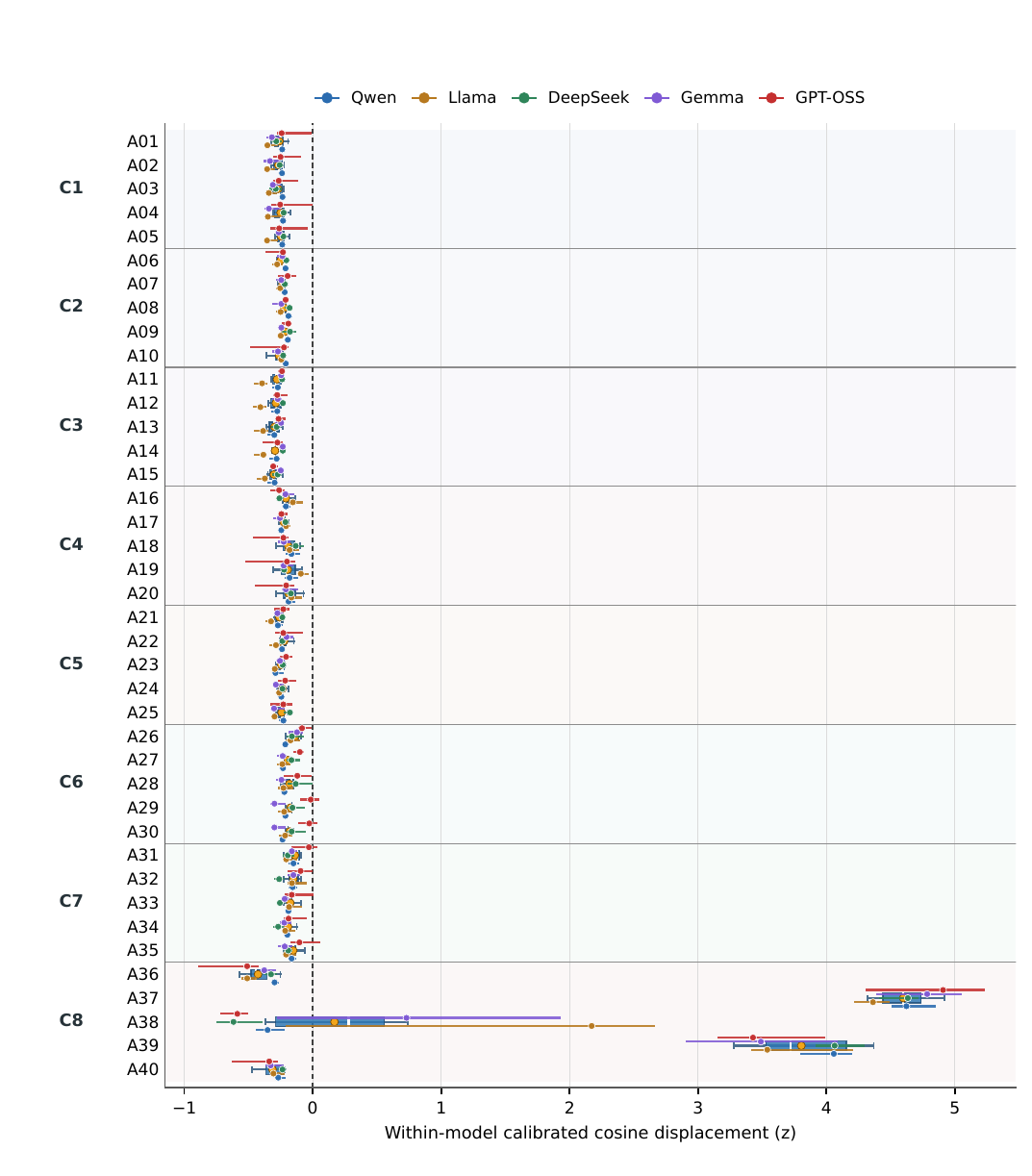}
  \caption{Target-specific displacement distributions for 40 attacks. Colored points and intervals show within-target medians and interquartile ranges; boxes summarize the corresponding cross-target profiles.}
  \label{fig:mechanism-displacement-distribution}
\end{figure*}

Figure~\ref{fig:mechanism-displacement-distribution} exposes the mechanism-level support behind the aggregate summaries: most attacks remain near the shared range, whereas A37--A39 produce the largest target-dependent departures and explain the C8 coverage challenge.

\subsection{Detector Adaptations and Ablations}

All detector adaptations use the same candidate records, target-specific calibration rows, held-out test rows, and nominal family-wise false-positive budget. Complete metrics underlying Figure~\ref{fig:detector-coverage-comparison} appear in Table~\ref{tab:detector-comparison-full}. PIShield-style selects a single layer on a disjoint development split and fits a robustly scaled linear probe, following PIShield's single-layer design~\cite{zou2025pishield}. Sentinel-style trains a 100-unit MLP probe at each layer, filters layers by development AUROC, removes probes with redundant attack-response profiles, and linearly fuses the selected layers, following Sentinel~\cite{liu2026sentinel}. Both operate on the frozen state sketches used by \audit and are paper-aligned reimplementations rather than official checkpoints.

The all-depth linear control flattens the same multidepth sketch before fitting one linear classifier. The text control fits a linear classifier to hashed word 1--2-grams and character 3--5-grams recovered through exact text hashes. These controls intentionally preserve strong average separation: their purpose is to test whether aggregate performance also provides mechanism coverage. All 240 LOAO and LOCO fold keys and their calibration, attack-test, and control-test counts match across the five methods.

\begin{table*}[!t]
  \centering
  \small
  \setlength{\tabcolsep}{4.6pt}
  \renewcommand{\arraystretch}{1.04}
  \caption{Detector comparison on identical candidate records, grouped splits, and target-specific calibration. TPR and FPR are percentages at the nominal 1\% family-wise budget. Min TPR is the lowest target-averaged recall among the held-out groups; coverage counts groups with at least 50\% recall.}
  \label{tab:detector-comparison-full}
  \begin{tabular}{@{}lrrrrrrr@{}}
    \toprule
    Detector & AUROC & AUPRC & TPR & FPR & Max FPR & Min TPR & Coverage \\
    \midrule
    \multicolumn{8}{c}{\emph{Panel A: leave-one-attack-out (40 mechanisms)}} \\
    \midrule
    \audit & 0.993 & 0.940 & 81.1 & 0.70 & 2.37 & 50.4 & 40/40 \\
    PIShield-style & 0.961 & 0.795 & 32.1 & 0.41 & 1.39 & 4.3 & 0/40 \\
    Sentinel-style & 0.956 & 0.773 & 28.1 & 0.23 & 0.78 & 4.5 & 0/40 \\
    All-depth linear & 0.992 & 0.968 & 90.7 & 0.47 & 1.59 & 0.5 & 38/40 \\
    Text n-gram & 0.989 & 0.954 & 92.7 & 0.71 & 2.39 & 0.0 & 37/40 \\
    \midrule
    \multicolumn{8}{c}{\emph{Panel B: leave-one-category-out (8 categories)}} \\
    \midrule
    \audit & 0.993 & 0.986 & 78.1 & 0.66 & 2.22 & 63.8 & 8/8 \\
    PIShield-style & 0.954 & 0.932 & 30.0 & 0.40 & 1.37 & 8.8 & 0/8 \\
    Sentinel-style & 0.948 & 0.922 & 26.0 & 0.33 & 1.10 & 8.9 & 0/8 \\
    All-depth linear & 0.983 & 0.983 & 87.2 & 0.46 & 1.54 & 41.3 & 7/8 \\
    Text n-gram & 0.991 & 0.988 & 91.5 & 0.16 & 0.55 & 41.4 & 7/8 \\
    \bottomrule
  \end{tabular}
\end{table*}

The component ablation in Table~\ref{tab:candidate-auditor-ablation} changes one design choice at a time while holding the evaluation protocol fixed. The oracle surface row is a nondeployable sensitivity analysis and is not part of the detector comparison.

\begin{table}[!t]
  \centering
  \small
  \setlength{\tabcolsep}{3.0pt}
  \renewcommand{\arraystretch}{1.06}
  \caption{Candidate-only LOAO ablation over 200 attack--model folds. The unified auditor combines ExtraTrees with a benign-only Mahalanobis gate under one calibrated false-positive budget; the oracle-routed row is nondeployable.}
  \label{tab:candidate-auditor-ablation}
  \vspace{3pt}
  \begin{tabular}{@{}lrrrr@{}}
    \toprule
    Features / detector & AUROC & TPR &
      FPR & Max FPR \\
    \midrule
    Projection / ET+MD & 0.993 & 81.1 & 0.70 & 2.37 \\
    Projection / ET & 0.993 & 78.4 & 0.45 & 1.52 \\
    Invariant / HistGB & 0.991 & 58.7 & 0.36 & 1.35 \\
    Invariant / ExtraTrees & 0.989 & 56.9 & \textbf{0.15} & \textbf{0.57} \\
    Invariant / Logistic & 0.981 & 42.3 & 0.26 & 0.98 \\
    Invariant / Mahalanobis & 0.945 & 15.1 & 0.88 & 3.30 \\
    Invariant energy & 0.927 & 2.2 & 0.32 & 1.21 \\
    Token length / Logistic & 0.771 & 0.0 & 0.53 & 1.98 \\
    \midrule
    Oracle surface label / ET+MD & 0.996 & 84.4 &
      0.52 & 1.76 \\
    \bottomrule
  \end{tabular}
\end{table}

\subsection{Protected-Outcome Holdout Values}

Complete protected-outcome results, including strict equal-cell metrics, deployment operating points, and record-pooled shortcut diagnostics, appear in Table~\ref{tab:core-audit-holdout}. These values underlie Figure~\ref{fig:protected-outcome-claim-boundary}.

\begin{table*}[!t]
  \centering
  \small
  \setlength{\tabcolsep}{3.2pt}
  \renewcommand{\arraystretch}{1.06}
  \caption{Protected-outcome prediction over 40,640 hash-joined executions. Macro metrics equal-weight the 194 evaluable attack--model cells; pooled AUC/AP are shortcut diagnostics and are not ranked. Within each holdout regime, boldface marks the best strict value per metric (higher AUROC, AP, and balanced accuracy; lower FPR95).}
  \label{tab:core-audit-holdout}
  \begin{tabularx}{\textwidth}{@{}L{0.130\textwidth}L{0.070\textwidth}C{0.060\textwidth}C{0.135\textwidth}C{0.070\textwidth}C{0.060\textwidth}C{0.060\textwidth}C{0.100\textwidth}X@{}}
    \toprule
    Feature set &
    Holdout &
    AUC cells &
    Macro AUROC (95\% CI) &
    Macro AP &
    Bal. acc. &
    FPR95 &
    Pooled AUC/AP &
    Audit use \\
    \midrule
    Token length &
    Within &
    194/200 &
    0.507 [0.495, 0.519] &
    0.423 &
    0.501 &
    0.966 &
    0.593 / 0.420 &
    Corpus cue; no alert signal. \\
    Token length &
    Joint &
    194/200 &
    0.502 [0.489, 0.514] &
    0.422 &
    0.500 &
    0.966 &
    0.438 / 0.335 &
    Fails transfer. \\
    Raw peak &
    Within &
    194/200 &
    0.503 [0.489, 0.518] &
    \textbf{0.425} &
    \textbf{0.503} &
    0.915 &
    0.649 / 0.464 &
    Composition shortcut. \\
    Raw peak &
    Joint &
    194/200 &
    0.511 [0.496, 0.525] &
    0.420 &
    \textbf{0.506} &
    0.912 &
    0.644 / 0.494 &
    Not an audit gate. \\
    Profile geometry &
    Within &
    194/200 &
    0.504 [0.490, 0.516] &
    0.424 &
    0.498 &
    0.911 &
    0.642 / 0.553 &
    Phenotype only. \\
    Profile geometry &
    Joint &
    194/200 &
    0.509 [0.495, 0.523] &
    0.423 &
    0.500 &
    0.915 &
    0.641 / 0.460 &
    Lacks outcome power. \\
    Geometry + shape &
    Within &
    194/200 &
    \textbf{0.512 [0.499, 0.525]} &
    \textbf{0.425} &
    0.499 &
    \textbf{0.902} &
    0.656 / 0.567 &
    Best within-model; abstain. \\
    Geometry + shape &
    Joint &
    194/200 &
    \textbf{0.518 [0.504, 0.531]} &
    \textbf{0.425} &
    0.500 &
    \textbf{0.896} &
    0.605 / 0.446 &
    Strongest strict probe; reject alert. \\
    \bottomrule
  \end{tabularx}
\end{table*}

\subsection{Extended Deployment Guidance and Limitations}

The deployment contract distinguishes unavailable prerequisites, constructor failure, refusal, successful compromise, and global utility failure. The same distinction applies to prompt construction, retrieval writes, agent actions, and parameter changes. Richer vectors, response-token trajectories, or source-conditioned states may recover information discarded by profile summaries, but any resulting detector still requires adaptive-evasion and distribution-shift evaluation. Least privilege, sandboxing, deterministic policy enforcement, and post-action verification remain necessary independently of the state auditor.

Five mechanisms per stratum cannot exhaust implementation diversity, and frozen snapshots provide reproducibility rather than timelessness. Normalized depth does not make layers functionally equivalent, and visible depth structure is not causal localization without intervention. The primary hybrid auditor uses one detector form across all 40 mechanisms, but recall remains nonuniform across mechanisms and targets. Operational use therefore requires a target-specific benign bank, workload-relevant hard negatives, and revalidation after serving changes. Candidate-state alerts also remain distinct from checkpoint or adapter integrity, which requires independently validated asset provenance and benign parameter-edit controls.

\section{Measurement Protocol and Trace Contract}
\label{app:measurement-protocol}

This appendix specifies the execution contract that turns the taxonomy into a measurement study.  It is intentionally more restrictive than a collection of successful attack examples: every attacked execution is paired with controls, bound to a source method and capability budget, assigned to a security track, and rejected when its provenance or trace integrity cannot be verified.

\subsection{Frozen Targets and Backends}

The Qwen3-4B, Llama-3.2-3B-Instruct, DeepSeek-R1-Distill-Qwen-7B, Gemma-3-12B-it, and GPT-OSS-20B snapshot revisions are \texttt{1cfa9a7}, \texttt{0cb88a4}, \texttt{916b56a}, \texttt{96b6f1e}, and \texttt{6cee5e8}, respectively. Generation uses pinned chat templates, BF16 or the model's native supported precision, and archived decoding parameters. The inference backend and judge are recorded per run. White-box collection uses a direct Transformers forward pass because the vLLM serving backend does not expose hidden states; this exception is recorded separately from the vLLM behavioral runs.

\subsection{Pre-Execution Gates}

For each attack mechanism, the source-artifact gate must record its primary publication URL, official code or model location, inspected revision, license, auxiliary models, supported target interfaces, expected access, and source budget. A smoke task assigns F3, F2, F1, or F0 fidelity using the definitions in Section~\ref{sec:attack-instantiation}. The following objects must be frozen before the reported analyses are run:

\begin{enumerate}[leftmargin=1.5em,itemsep=2pt]
\item the 40-mechanism registry and alternate log;
\item exact checkpoint, tokenizer, chat-template, and inference revisions;
\item base-task groups, security predicates, hard negatives, and split assignments;
\item attack budgets and stopping rules for each mechanism;
\item trace event anchors, layer grid, feature definitions, and rejection rules; and
\item outcome rubrics, judge prompts, adjudication policy, and statistical analysis plan.
\end{enumerate}

No outcome from the final test partition may be used to revise these objects. Necessary corrections receive a versioned protocol amendment and invalidate affected final runs.

\begin{figure}[!t]
  \centering
  \begin{tikzpicture}[
  font=\small,
  card/.style={
    draw=neutral,
    line width=0.65pt,
    rounded corners=2pt,
    align=center,
    inner xsep=4pt,
    inner ysep=5pt
  },
  arrow/.style={
    draw=neutral,
    line width=0.75pt,
    -{Latex[length=1.8mm,width=1.3mm]}
  }
]

\node[card, fill=cTwo!9, text width=3.45cm] (inputs) at (0,0)
  {\textbf{Attack inputs}\\method, budget, access, F0--F3};

\node[card, fill=cThree!9, text width=3.45cm] (controls) at (4.05,0)
  {\textbf{Paired controls}\\clean, attack, carrier, counterfactual};

\node[card, fill=cSix!9, text width=5.25cm] (replay) at (2.025,-1.55)
  {\textbf{Pinned replay}\\five model families; deterministic and hash-checked};

\node[card, fill=cSeven!9, text width=3.45cm] (traces) at (0,-3.15)
  {\textbf{White-box traces}\\states, logits, depth, provenance};

\node[card, fill=cFive!9, text width=3.45cm] (behavior) at (4.05,-3.15)
  {\textbf{Behavior records}\\outcome, cost, failures, reason codes};

\node[card, fill=analysisblue!9, text width=7.35cm] (matrix) at (2.025,-4.75)
  {\textbf{Frozen evaluation matrix}\\paired IDs, held-out splits, deduplication, checksums};

\node[card, fill=cEight!9, text width=7.35cm] (measurements) at (2.025,-6.25)
  {\textbf{Audit measurements}\\shift, transfer, transition, and outcome probes};

\draw[arrow] (inputs.south) -- (replay.north west);
\draw[arrow] (controls.south) -- (replay.north east);
\draw[arrow] (replay.south west) -- (traces.north);
\draw[arrow] (replay.south east) -- (behavior.north);
\draw[arrow] (traces.south) -- (matrix.north west);
\draw[arrow] (behavior.south) -- (matrix.north east);
\draw[arrow] (matrix.south) -- (measurements.north);

\end{tikzpicture}
  \caption{Measurement pipeline. Frozen attack inputs are executed on pinned target snapshots. Behavioral records and paired internal profiles are produced independently and joined only through exact input hashes; derived results are rebuilt from checksummed artifacts.}
  \label{fig:measurement-pipeline}
\end{figure}
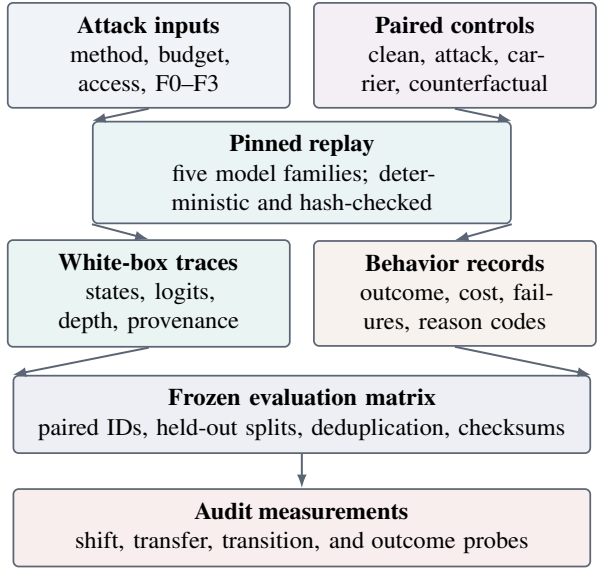

\subsection{Protected-Execution Record}

Every protected execution emits one immutable manifest row and one trace object.  Table~\ref{tab:execution-schema} lists the minimum fields.  Raw prompts and outputs may reside in a restricted partition, but their hashes and structural metadata remain in the shareable aggregate.

\begin{table*}[!t]
  \centering
  \small
  \caption{Minimum execution and trace schema.}
  \label{tab:execution-schema}
  \begin{tabularx}{\textwidth}{L{0.21\textwidth}Y Y}
    \toprule
    \textbf{Field group} & \textbf{Required fields} & \textbf{Integrity rule} \\
    \midrule
    Identity
      & run, parent task, paper, method, category, condition, split, seed
      & descendants inherit the parent-task split \\
    Model
      & checkpoint, weight and tokenizer hashes, template, precision, backend
      & every hash matches the frozen model manifest \\
    Attack
      & constructor revision, access, budget, calls, candidate and state hashes
      & calls do not exceed the source-method budget \\
    Security
      & track, protected event, deterministic predicate, judge version
      & numerator and denominator are track-specific \\
    Provenance
      & token, turn, channel, trust, record, producer, write, parameter region
      & every protected-event input maps to a provenance unit \\
    Trace
      & layer states, logits, entropy, optional attention, event indices
      & instrumentation preserves deterministic output tokens \\
    Outcome
      & predicate result, judge labels, disagreement, adjudication, validity
      & ambiguous or invalid runs are not silently treated as failures \\
    \bottomrule
  \end{tabularx}
\end{table*}

\subsection{Condition Groups}

The base task is the grouping and resampling unit.  A direct-input group contains a clean request, an attacked request, a length- and carrier-matched benign request, and an ordinary-channel intent control.  A dialogue group also contains history reset, matched benign escalation, and final-turn-only conditions.  An indirect-injection group contains source-present, source-removed, channel-transplanted, and relevance-matched benign-source conditions.  A persistent-state group contains clean state, poison-present state, relevance-matched insertion, a fresh-session read, and a state-removal condition.  A parameter group crosses clean versus modified checkpoint with trigger absent versus present.

Where a source method can produce valid successful and unsuccessful candidates, both are sampled under the same base task and budget.  Invalid syntax, failed retrieval, unavailable protected events, and trace corruption receive distinct validity codes.  The main behavioral denominator contains eligible protected executions, while end-to-end yield additionally includes construction and system failures.  This separation prevents implementation fragility from being mistaken for model robustness.

\subsection{Outcome Adjudication}

Each security track has a dedicated outcome: harmful compliance requires both substantive assistance and prohibited content; instruction integrity requires following the trusted objective rather than an untrusted instruction; retrieval or memory integrity requires an attacker-targeted belief or use of poisoned state; unauthorized action requires a concrete action proposal or call outside the task policy; and backdoor behavior requires a trigger-conditioned effect relative to both clean model and trigger-absent controls.  No global success ratio pools these denominators.

Deterministic predicates are used whenever the environment exposes an authorization, retrieval, target-string, or action state. For any claim that relies on semantic outcome labels, the evidence package must document independent judgments by two frozen judges and a blinded human audit stratified by track, method, and judge disagreement. That validation must report agreement, false-positive and false-negative estimates, ambiguous rate, and sensitivity to counting ambiguous cases either way. Until those artifacts are present, automated labels are treated as measurement instruments rather than validated semantic ground truth.

\subsection{Internal Trace Processing}

Residual states are collected at the final input token and at protected decoding events, then mapped to normalized depth.  Within-model whitening and clean-reference statistics are fitted on the discovery training partition only.  Source-conditioned summaries aggregate by provenance unit before any cross-model feature adapter is trained.  Raw attention is stored where available for diagnostic comparison but never serves as the sole causal evidence.

A trace fails quality control if model or prompt hashes disagree, hook-enabled deterministic decoding changes tokens, a required layer or event is missing, the provenance map is incomplete, non-finite values occur, or the protected event falls outside the capture interval. Rejected traces remain in the run ledger with their reason. Exclusion rates are reported by category, mechanism, model, task, and condition.

\subsection{Statistical Analysis}

The descriptive unit is a protected execution, but the inferential unit is the attack mechanism and base task. Within-category estimates first aggregate descendants to their parent task and then mechanisms to a category. Inferential category contrasts require equal-category macro weighting and hierarchical bootstrap intervals that resample mechanisms within category and tasks within mechanism. Any mixed-effects analysis must include condition and the frozen factor vector, with random effects for mechanism, model, and base task. Model comparisons must not treat thousands of generated prompts as independent attack methods.

Any layer-wise hypothesis test must use false-discovery-rate control, and a discovery finding must repeat in a held-out confirmation partition with a direction frozen before the reported analyses. A cross-mechanism state signature additionally requires cross-category coverage, mechanism-level consistency, model holdout, hard-negative specificity, and at least one aligned intervention. Any label-free exploratory analysis must be completed before revealing the primary categories. Category coherence, alternative-label sensitivity, leave-one-mechanism sensitivity, and factor-vector models determine how much explanatory weight the taxonomy can support.

\section{\audit Development and Evaluation Protocol}
\label{app:audit-protocol}

\begin{figure*}[!t]
  \centering
  \small
  \begin{tikzpicture}[
    font=\small,
    x=1cm,y=1cm,
    box/.style={draw=#1!70!black, fill=#1!10, rounded corners=2pt,
      line width=0.55pt, align=left, inner sep=4.2pt},
    badge/.style={draw=#1!75!black, fill=#1!20, rounded corners=2pt,
      line width=0.7pt, align=center, inner sep=3.2pt, font=\bfseries\small},
    gate/.style={-Latex, line width=0.7pt, draw=neutral!85},
    fail/.style={-Latex, line width=0.8pt, draw=red!70!black},
    metric/.style={font=\bfseries\small}
  ]
    \def\Yzero{0}
    \def\xone{0}
    \def\xtwo{3.15}
    \def\xthree{7.05}
    \def\xfour{11.35}
    \def\xfive{15.2}

    \node[font=\bfseries, anchor=west] at (\xone,0.25) {Observable trace};
    \node[font=\bfseries, anchor=west] at (\xtwo,0.25) {Frozen evidence};
    \node[font=\bfseries, anchor=west] at (\xthree,0.25) {Claim promoted};
    \node[font=\bfseries, anchor=west] at (\xfour,0.25) {\audit action};

    \node[badge=analysisblue, minimum width=2.6cm, minimum height=0.72cm]
      (b0) at (\xone,-1.0) {L0\\[-1pt]Behavioral matrix};
    \node[box=analysisblue, text width=3.25cm, minimum height=0.88cm, anchor=west]
      (e0) at (\xtwo,-1.0) {\textbf{192.5k} judged\\ASR: \textbf{18.3--48.7\%}\\refusal: \textbf{1.6--25.1\%}};
    \node[box=analysisblue, text width=3.72cm, minimum height=0.88cm, anchor=west]
      (c0) at (\xthree,-1.0) {Effectiveness is target-model dependent; no global success summary.};
    \node[box=analysisblue, text width=3.25cm, minimum height=0.88cm, anchor=west]
      (a0) at (\xfour,-1.0) {Report denominators and exposure};

    \node[badge=cSix, minimum width=2.6cm, minimum height=0.72cm]
      (b1) at (\xone,-2.50) {L1\\[-1pt]Paired state shift};
    \node[box=cSix, text width=3.25cm, minimum height=0.88cm, anchor=west]
      (e1) at (\xtwo,-2.50) {\textbf{100k} matched pairs\\cosine + relative-$\ell_2$\\depth profiles};
    \node[box=cSix, text width=3.72cm, minimum height=0.88cm, anchor=west]
      (c1) at (\xthree,-2.50) {Attacks leave structured calibrated internal phenotypes.};
    \node[box=cSix, text width=3.25cm, minimum height=0.88cm, anchor=west]
      (a1) at (\xfour,-2.50) {Emit descriptive trace};

    \node[badge=cSeven, minimum width=2.6cm, minimum height=0.72cm]
      (b2) at (\xone,-4.00) {L2\\[-1pt]Mechanism fingerprint};
    \node[box=cSeven, text width=3.25cm, minimum height=0.88cm, anchor=west]
      (e2) at (\xtwo,-4.00) {LOAO C1--C8 recovery\\state: \textbf{82.5\%}\\behavior+state: \textbf{85.0\%}};
    \node[box=cSeven, text width=3.72cm, minimum height=0.88cm, anchor=west]
      (c2) at (\xthree,-4.00) {Profile geometry supports mechanism-level triage.};
    \node[box=cSeven, text width=3.25cm, minimum height=0.88cm, anchor=west]
      (a2) at (\xfour,-4.00) {Promote mechanism evidence};

    \node[badge=cEight, minimum width=2.6cm, minimum height=0.72cm]
      (b3) at (\xone,-5.85) {L3\\[-1pt]Reference auditor};
    \node[box=cEight, text width=3.25cm, minimum height=0.88cm, anchor=west]
      (e3) at (\xtwo,-5.85) {projection/ET+MD LOAO\\AUROC: \textbf{0.993}\\TPR/FPR: \textbf{81.1/0.70\%}};
    \node[box=cEight, text width=3.72cm, minimum height=0.88cm, anchor=west]
      (c3) at (\xthree,-5.85) {Target-calibrated benign references detect attack-like perturbations.};
    \node[box=cEight, text width=3.25cm, minimum height=0.88cm, anchor=west]
      (a3) at (\xfour,-5.85) {Emit attack-evidence alert};

    \node[badge=red, minimum width=2.6cm, minimum height=0.72cm]
      (b4) at (\xone,-7.70) {L4\\[-1pt]Outcome success};
    \node[box=red, text width=3.25cm, minimum height=0.88cm, anchor=west]
      (e4) at (\xtwo,-7.70) {strict success probe\\AUROC: \textbf{0.518}\\FPR95: \textbf{0.896}};
    \node[box=red, text width=3.72cm, minimum height=0.88cm, anchor=west]
      (c4) at (\xthree,-7.70) {Current features do not validate a binary success alarm.};
    \node[box=red, text width=3.25cm, minimum height=0.88cm, anchor=west]
      (a4) at (\xfour,-7.70) {\textbf{Abstain}; require outcome validation};

    \foreach \i in {0,1,2,3,4}{
      \draw[gate] (b\i.east) -- (e\i.west);
      \draw[gate] (e\i.east) -- (c\i.west);
      \draw[gate] (c\i.east) -- (a\i.west);
    }

    \draw[gate] (b0.south) -- (b1.north);
    \draw[gate] (b1.south) -- (b2.north);
    \draw[gate] (b2.south) -- (b3.north);
    \draw[fail] (b3.south) -- (b4.north);
    \node[anchor=west, text=red!70!black, font=\bfseries\footnotesize] at (1.08,-6.77)
      {claim boundary};
  \end{tikzpicture}
  \caption{\audit evidence ladder induced by the 40-mechanism study. The visual contract shows how observed traces are promoted into audit claims: behavioral coverage supports exposure reporting, paired state shifts support descriptive traces, calibrated profile geometry supports mechanism-level triage, and target-calibrated benign references support attack-like perturbation auditing. The final gate fails only for protected-outcome success prediction: \audit reports attack evidence while abstaining from claims that the protected violation succeeded.}
  \label{fig:core-audit-claim-ladder}
\end{figure*}
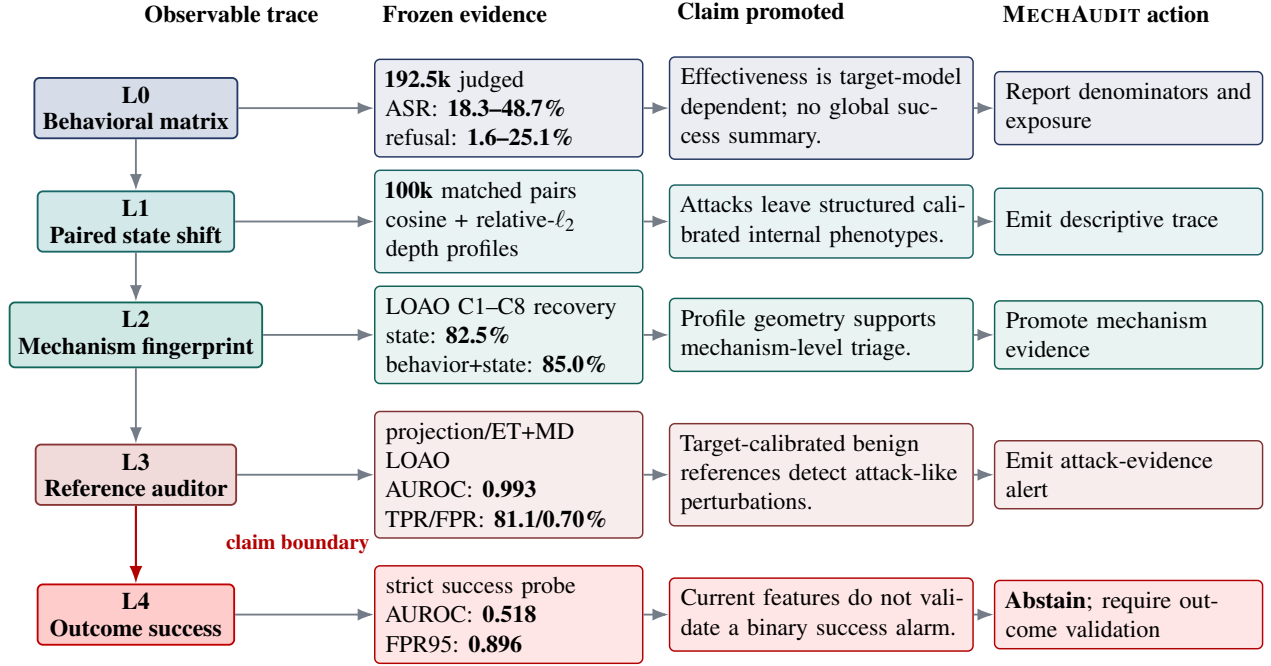

Figure~\ref{fig:core-audit-claim-ladder} summarizes the evidence contract used throughout this appendix: each higher claim requires an additional frozen test, and the auditor abstains when the evidence supports attack exposure but not protected-outcome success.

\subsection{Frozen Data Lifecycle}

The implemented audit pipeline defines three disjoint data stages. The offline discovery stage constructs matched clean--attack trajectories and freezes the multidepth state representation. The auditor-development stage freezes one random projection shared across all targets, then fits a separate ExtraTrees head, benign-only Mahalanobis gate, reference bank, robust target scaler, and pair of operating thresholds for each target model. The final stage contains untouched base-task groups under mechanism, category, and model holdouts fixed before scoring. The split audit checks base IDs, candidate-input hashes, model-specific feature hashes, and cross-label feature hashes; all 310 evaluated folds contain zero detected overlaps.

Category and attack labels define grouped holdouts but are never supplied to the candidate-only scorer. Likewise, the primary scorer does not receive an intervention-surface label. This constraint is important because a surface label inferred from the evaluated attack would reveal the held-out mechanism family. The reported primary result therefore uses only the frozen projection and the ExtraTrees and Mahalanobis gates calibrated for the declared target model. Parameter or training provenance is retained as an audit boundary rather than used as an oracle routing signal.

\subsection{Fast and Confirmatory Paths}

The deployment-facing path consumes one instrumented candidate execution. It extracts the frozen multidepth state vector, applies the shared projection, scales the coordinates against the target's benign bank, and evaluates that target's ExtraTrees and Mahalanobis gates. Scoring receives no matched clean execution, attack identity, category, intervention surface, or protected-outcome label. An alert is the union of the two gate decisions under one split false-positive budget. Its output is attack-exposure evidence with a threshold and calibration record; it is neither a mechanism attribution nor proof that the protected violation succeeded.

The matched clean--attack path is an offline diagnostic used to construct the mechanism atlas and test category recovery. Because it has an oracle clean counterpart, it is not reported as a deployment detector. The current artifact also makes no causal source or layer claim: activation editing and bidirectional patching remain future confirmation tests rather than components of the deployed scorer.

\paragraph{Standalone reference banks.}
The reported deployment bank is constructed from target-specific clean executions and hard-negative families. Clean members of the paired atlas may enter as ordinary benign reference rows, but their pairing is never exposed to the scorer and no candidate is subtracted from its counterpart at test time. Model identity, checkpoint revision, tokenizer behavior, context configuration, layer geometry, and bank construction policy are bound to the evaluation manifest; a new target or serving revision requires a new benign bank and thresholds.

The ExtraTrees head uses 400 trees, balanced class weights, square-root feature subsampling, and a minimum of eight samples per leaf. The Mahalanobis gate uses the seven 21-bin invariant profiles, median centering, 10th--90th percentile scaling, clipping to $[-20,20]$, and shrinkage covariance fitted only on benign development rows. For target $m$, gate $g$, and nominal union false-positive budget $\alpha$, the threshold is
\begin{equation}
  \tau_{m,g}=\max_k Q_{1-\alpha/2}\!\left(\{s_g(b):b\in\mathcal{B}_{m,k}\}\right),
  \label{eq:gate-threshold}
\end{equation}
where $\mathcal{B}_{m,k}$ is benign control family $k$ for target $m$. We use $\alpha\in\{0.01,0.001\}$, assign half of the budget to each gate, and alert when either gate exceeds its threshold. The maximum over control families prevents a large benign family from hiding a smaller high-scoring family. No scaler, gate, or threshold is tuned per attack mechanism.

\subsection{Evaluation Matrix}

\begin{table*}[!t]
  \centering
  \small
  \caption{Audit regimes fixed before the reported analyses and the shortcut each regime removes.}
  \label{tab:audit-regimes}
  \begin{tabularx}{\textwidth}{L{0.19\textwidth}L{0.28\textwidth}Y}
    \toprule
    \textbf{Regime} & \textbf{Held-out unit} & \textbf{Question} \\
    \midrule
    Interpolation & base-task groups & Can the implementation fit represented mechanisms? \\
    Unseen mechanism & one complete attack mechanism & Does it transfer beyond the represented attacks? \\
    Unseen category & one whole mechanism stratum & Does a shared execution signature remain useful? \\
    Unseen model & one complete target model & Does the normalized representation transfer? \\
    Unseen channel & one delivery channel & Does provenance generalize beyond a known transport? \\
    Joint & attack mechanisms and a complete target model & Does improvement survive both mechanism and model novelty? \\
    Adaptive & final monitor-aware candidates & What risk remains under the source capability budget? \\
    \bottomrule
  \end{tabularx}
\end{table*}

The audit regimes in Table~\ref{tab:audit-regimes} define an evaluation contract rather than guaranteeing a completed result. A completed regime must report per-mechanism results, equal-category macro average, worst category, per-model results, hard-negative false positives, calibration, and a hierarchical interval. Leave-one-category training has no expert for the held-out category and permits mechanism abstention. The principal binary risk metric remains mandatory. Unseen-model evaluation must report benign-only target calibration separately from strict architecture-statistic transfer.

\subsection{Metrics and Thresholds}

For a completed detection evaluation, the primary metrics are TPR at 1\% FPR and AUPRC, paired at the mechanism level against the strongest baseline chosen on development data. TPR at 0.1\% FPR may be reported only when the benign test count supports that empirical resolution. Deployment-oriented claims additionally require false positives per 10,000 benign executions, deployment-prevalence precision, Brier score, expected calibration error, and selective risk. Category attribution uses macro-F1 and abstention-aware accuracy. A localization claim requires AUPRC, recall@$k$, and span IoU where applicable; depth localization additionally requires interval IoU and absolute normalized-depth error. Metrics not supported by the available test resolution must be marked unreported rather than inferred.

The 0.1\% and 1\% operating thresholds are fitted on an independent calibration fold and frozen before final attack traces are scored.  If a subgroup lacks enough benign examples to estimate low FPR, the exact resolution is stated rather than obtained by smoothing or extrapolation.  No final threshold is selected from test ROC curves.

\subsection{Causal Faithfulness and Adaptive Evaluation}

A causal-localization claim requires edits at the auditor's top source and depth to be compared with uniformly random, length-matched, provenance-matched, attention-selected, and semantic controls. The corresponding analysis must report outcome-margin change, bidirectional patching consistency, benign semantic change, and the fraction of executions exceeding each control. A causal label is defined by these interventions, not by the attack author's prompt span alone.

An adaptive-robustness claim requires monitor-aware construction that preserves each source method's access, calls, state writes, training budget, and validity constraints. The fixed final auditor must not be updated after seeing these candidates. Such an evaluation must separate ordinary attack success, success against the auditor, evasion-conditioned success, cost inflation, surface change, hard-negative false-alarm pressure, and uncertainty. Any second round of adversarial training must use a new partition and be labeled as a separate development result. The reported non-adaptive evaluation does not by itself establish adaptive robustness.

\subsection{Baseline and Ablation Contract}

Every reported baseline must receive the same parent-task splits and the information its published interface permits. External auditing methods may be included only when their source artifacts and target interface support a faithful execution; adaptations must be distinguished from official checkpoints. Table~\ref{tab:detector-comparison-full} contains the completed same-record comparison with single-layer, complementary-layer, flat multi-layer, and lexical detectors. Comparisons requiring unavailable inputs, such as output trajectories or attention tensors, remain evidence requirements rather than implicit results.

The completed ablation compares token-length, invariant-energy, Mahalanobis, logistic, ExtraTrees, and histogram-gradient-boosting variants with the same candidate-only inputs and benign calibration protocol. Components receive credit only when a reported grouped holdout supports the gain. Category experts, a learned localizer, transition features, and an oracle surface route are not part of the primary implementation; the oracle route is retained only as a labeled nondeployable sensitivity analysis.

\end{document}